\documentclass{article}
\usepackage{arxiv}

\usepackage{epstopdf}% To incorporate .eps illustrations using PDFLaTeX, etc.
\usepackage[caption=false]{subfig}

\usepackage[natbibapa,nodoi]{apacite}
\usepackage{xcolor}
\usepackage[utf8]{inputenc} % allow utf-8 input
\usepackage{mathptmx,amsmath,amssymb,xfrac,mathtools}
\usepackage{textcomp,marvosym}
\usepackage{fixltx2e}
\usepackage{nameref}
\usepackage{microtype}      % microtypography
\usepackage{esdiff,booktabs,longtable}
\usepackage{lastpage,fancyhdr}
\usepackage[T1]{fontenc}    % use 8-bit T1 fonts
\usepackage{url}            % simple URL typesetting
\usepackage{booktabs}       % professional-quality tables
\usepackage{amsfonts}       % blackboard math symbols
\usepackage{nicefrac}       % compact symbols for 1/2, etc.
\usepackage{lipsum}
\usepackage{graphicx}
\usepackage{amsmath}
\usepackage{mathtools}
\usepackage[inline]{enumitem}
\usepackage{array}
\usepackage{multirow}
\usepackage{siunitx}
\usepackage{blindtext}
\usepackage{hhline}
\usepackage{epstopdf}
\usepackage{tabularx}
\usepackage[aboveskip=1pt,labelfont=bf,labelsep=period,justification=centering,singlelinecheck=off]{caption}
\usepackage{bm}

\newcolumntype{+}{!{\vrule width 2pt}}
\newcolumntype{b}{X}
\newcolumntype{s}{>{\hsize=.5\hsize}X}

\let\svthefootnote\thefootnote
\newcommand\blankfootnote[1]{%
  \let\thefootnote\relax\footnotetext{#1}%
  \let\thefootnote\svthefootnote%
}

\setlist[enumerate,1]{%
  label=\arabic*.,
}
\setlist*[enumerate,1]{%
  label=(\roman*),
}

\newlist{legal}{enumerate}{10}
\setlist[legal]{label*=\arabic*.}

\graphicspath{ {./images/} }

\newenvironment{conditions*}
  {\par\vspace{\abovedisplayskip}\noindent
   \tabularx{\columnwidth}{>{$}l<{$} @{${}\quad{}$} >{\raggedright\arraybackslash}X}}
  {\endtabularx\par\vspace{\belowdisplayskip}}

\begin{document}
\vspace*{0.2in}

\begin{flushleft}
{\Large
\textbf\newline{Data-driven Fair Resource Allocation For Novel Emerging Epidemics: A COVID-19 Convalescent Plasma Case Study}
}
\newline
\\
Maryam Akbari-Moghaddam\textsuperscript{1,*},
Na Li\textsuperscript{1,2,3},
Douglas G. Down\textsuperscript{1},
Donald M. Arnold\textsuperscript{3,4},
Jeannie Callum\textsuperscript{5,6},
Philippe Bégin\textsuperscript{7,8},
Nancy M. Heddle \textsuperscript{3,4}.
\\
\bigskip
\textbf{1} Department of Computing and Software, McMaster University, Hamilton, Ontario L8S 4L7, Canada
\\
\textbf{2} 
Community Health Sciences, University of Calgary, Calgary, Alberta T2N 1N4, Canada 
\\
\textbf{3} McMaster Centre for Transfusion Research, Department of Medicine, McMaster University, Hamilton, Ontario L8N 3Z5, Canada
\\
\textbf{4} Department of Medicine, McMaster University, Hamilton, Ontario L8N 3Z5, Canada
\\
\textbf{5} Department of Pathology and Molecular Medicine, Kingston Health Sciences Centre, Kingston, Ontario K7L 2V7, Canada
\\
\textbf{6} Department of Pathology and Molecular Medicine, Queen's University, Kingston, Ontario K7L 3N6, Canada
\\
\textbf{7} Section of Allergy, Immunology and Rheumatology, Department of Pediatrics, CHU Sainte-Justine, Montreal, Quebec H3T 1C5, Canada
\\
\textbf{8} Department of Medicine, CHUM, Université de Montréal, Montreal H3T 1J4, Quebec, Canada
\\

\bigskip

* Corresponding author \\
Email: akbarimm@mcmaster.ca

\end{flushleft}
\begin{abstract}

Epidemics are a serious public health threat, and the resources for mitigating their effects are typically limited. Decision-makers face challenges in forecasting the supply and demand for these resources as prior information about the disease is often not available, the behaviour of the disease can periodically change (either naturally or as a result of public health policies) and can differ by geographical region. Randomized controlled trials (RCTs) using scarce resources such as blood products as a randomized intervention are affected by epidemics. In this work, we discuss a model that is suitable for short-term real-time supply and demand forecasting during emerging outbreaks. We consider a case study of demand forecasting and allocating scarce quantities of COVID-19 Convalescent Plasma (CCP) in an international multi-site RCT involving multiple hospital hubs across Canada (excluding Québec). We propose a data-driven mixed-integer programming (MIP) resource allocation model that assigns available resources to maximize a notion of fairness among the resource-demanding entities. Numerical results from applying our MIP model to the case study suggest that our approach can help balance the supply and demand of limited products such as CCP and minimize the unmet demand ratios of the demand entities. We analyze the sensitivity of our model to different allocation settings and show that our model assigns equitable allocations across the entities.
\end{abstract}

\section{Introduction} \label{Introduction}

Epidemics \blankfootnote{This paper has been submitted to the INFOR: Information Systems and Operational Research journal.} have impacted the world many times, and will occur again in the future. Emergency responses to these epidemics can either be pre-event or post-event, where here event refers to the initiation of an infectious disease \citep{dasaklis2012epidemics}. Forecasting the potential dangers and planning the necessary steps to deal with an epidemic are considered as pre-event tasks. Post-event responses occur after the disease has started spreading and is still in progress. The corresponding actions at these points are associated with treatment and allocating the corresponding available resources. Our focus in this paper is on post-event response situations.

Resource allocation decisions during emerging epidemics are challenging due to several key reasons.
First, limited knowledge and historical data about disease demographics make it difficult to predict the demand for particular resources. Second, because epidemics are usually unexpected and can spread rapidly, there are often limited health care resources (vaccines, blood products, medical equipment, etc.) compared to the total number of entities requesting them. Determining how to fairly allocate the limited resources becomes a challenge. Third, the demand can vary significantly between geographically dispersed entities. Finally, the decisions must be made in a timely manner. These issues motivate the investigation of supply and demand forecasting models for scenarios where there are small amounts of available data and the data can exhibit fundamental changes in behaviour. It is of interest to incorporate these models into algorithms that yield fair allocations.

Clinical trials can have different durations depending on the underlying research, with many lasting as short as several months \citep{piantadosi2017clinical}. Some of these clinical trials may require rapid roll-out to different participating sites right after the planning phase, which might not allow for prior data collection by the sites. Randomized Controlled Trials (RCTs) can face additional limitations in data collection within the study duration as different experimental interventions can be associated with specific time marks. During uncertain situations like emerging epidemics, RCTs that use scarce resources such as blood products as a randomized intervention can be affected when there is a need for a timely and fair resource allocation \citep{stawicki20202019}. Convalescent plasma is a blood product which has been used as a potential treatment for a number of diseases such as Ebola \citep{kraft2015use, van2016evaluation}, influenza \citep{hung2011convalescent, zhou2007treatment}, and COVID-19 \citep{chen2020convalescent}. COVID-19 Convalescent Plasma (CCP) contains antibodies, or special proteins, generated by the body’s immune system in response to the novel coronavirus \citep{chen2020convalescent}. It has been considered as an experimental treatment for hospitalized COVID-19 patients in a number of RCTs worldwide \citep{begin2021convalescent,li2020effect,simonovich2021randomized,gharbharan2020convalescent}. Despite the importance of blood supply and demand management for scarce products in RCTs, there is a lack of published research that provides insights on the problem. This can be related to several challenges encountered when making decisions on CCP allocation in an RCT setting. When such a scarce blood product serves as a randomized intervention in an RCT, poor linkage across blood donor recruitment, inventory management and the randomization process can raise biases \citep{li2022data}.

In this work, we tackle the real-time allocation of scarce resources during epidemics to entities located in widespread geographical locations and with different resource requirements. Furthermore, we consider an RCT setting which requires rapid roll-out and has limited historical data due to its short study duration. We are interested in demand forecasting models that can predict short-term demand in real-time. The demand forecasts directly impact the resource allocation decisions as inaccurate forecasts may lead to inefficient and unfair use of limited (and valuable) resources. We note that there are different notions of fairness (balance) in terms of resource allocation in the literature \citep {kumar2000fairness}. In our case, we define fairness as minimization of the entities' unmet demand ratios, but one could also generalize fairness to other notions. For instance, 
% Karsu et al.\ 
\cite{karsu2020balance} propose an approach where imbalance is defined as the deviation from a reference distribution determined by the decision-maker. They show that in resource allocation problems, it is possible to maintain a mixed-integer programming (MIP) structure even after generalizing the notion of fairness. In general, there are wide-ranging views of fairness and how fairness metrics can be incorporated into optimization problems \citep[see][]{karsu2014incorporating, smith2013bicriteria, mestre2012organizing, heitmann2014preference, karsu2015inequity}. We evaluate our proposed model on a case study of demand forecasting and CCP distribution in a multi-site RCT, where \begin{enumerate*}\item there were limited historical supply and demand data, \item the supply was limited and restricted by manufacturing policies and donor recruitment,\item the demand arising from the demand entities was heterogeneous, and \item specific clinical requirements were needed for administrating CCP transfusion. More details of the RCT, including how the approach presented in this paper is embedded and the resulting observations from a clinical perspective, can be found in \citep {li2022data}. \end{enumerate*}

We make the following contributions: First, we discuss real-time short-term forecasting of supply and demand of scarce resources in epidemics with limited historical data. We propose the use of a forecasting model that does not require indeterminate epidemiological parameters (such as location and time-specific parameters) and thus does not require periodically updating the parameters. Secondly, we address challenges that may arise in an online setting due to extrapolation and sparse data. Next, we propose a data-driven MIP model for real-time multi-location allocation of scarce resources regularly and fairly to entities, which have heterogeneous demand. This approach maximizes a notion of fairness among the resource-demanding entities. Finally, numerical results of applying our model in a CCP case study show that our approach yields fair allocations that are both close to the scenario 
where supply and demand are known (rather than forecast) and are preferable to what was used in practice.

The rest of the paper is organized as follows. Section \ref{LiteratureReview} presents the existing literature on demand forecasting and resource allocation approaches during infectious disease outbreaks and our motivation for this work. We describe our data-driven resource allocation problem in Section \ref{problemdescription}. Section \ref{supplyanddeamndforecasting} discusses in depth the supply and demand forecasting methods that we use and we define our proposed MIP resource allocation model in Section \ref{optimization}. The CCP case study, the numerical results of applying our MIP model to the case study, and sensitivity analysis of the model are discussed in Section \ref{SimulationResultsplusInsights}. We conclude this work and discuss how it may inform responses to future pandemics in Section \ref{Conclusion}.

\section{Motivation and Related Work} 
\label{LiteratureReview}

There has been a variety of work in various fields that tackle the problem of resource allocation during infectious disease outbreaks. In what follows, we divide the body of work into six categories and discuss the motivation behind our proposed methodology.
\hfill \\ \break
\textbf{Compartmental Models and Resource Allocation}

Epidemiological compartmental models, consisting of a set of nonlinear ordinary differential equations, can help model the dynamics of different epidemiological variables during a pandemic \citep{brauer2008compartmental}. These models can give insight into disease-related information such as spread rate, the duration of an epidemic, and the total number of infected and recovered patients. Decision-makers can employ compartmental models to derive demand for medical resources to guide resource allocation decisions. Focusing on the COVID-19 outbreak \citep{velavan2020covid}, there have been many applications and tools developed by different organizations worldwide to 
forecast infections, hospitalizations, and deaths using compartmental models \citep{tomar2020prediction, gong2020tool, chowdhury2021early}. For 
instance, CHIME \citep {weissman2020locally} is a tool based on a Susceptible-Infectious-Recovered (SIR) model that 
can be used for forecasting the number of daily hospitalized COVID-19 patients in the short-term (e.g., up to $30$ days).

\hfill \\ 
\textbf{Demand Forecasting Models and Resource Allocation}

Another approach that researchers have studied for forecasting healthcare resources is using time series models or machine learning methods. The references for this approach are extensive, thus we only discuss a few studies as examples. 
% Ferstad et al.\
\cite{ferstad2020model} introduce a time series model to forecast the availability and utilization of intensive and acute care beds. 
% Nikolopoulos et al.\
\cite{nikolopoulos2021forecasting} use epidemiological and deep learning models to forecast the excess demand for products and services considering auxiliary data and simulating governmental decisions, while 
% Li et al.\ 
\cite{li2021decision} combine ideas from statistical time series modelling and machine learning to develop a hybrid demand forecasting model for red blood cell components using clinical predictors.

\hfill \\ 
\textbf{Segmented Regression Models and Resource Allocation}

Piecewise Linear Regression (PLR), also known as segmented linear regression, forecasting models are a special case of a larger set of models known as spline functions \citep{suits1978spline}. Modelling the regression function in "pieces" can be helpful when dealing with sparse data because we can still use linear regression models for data that does not fit a single line. To be more specific, PLR is a simple model that makes understanding the data easier by solving several linear regressions. Points at which the behaviour changes are called breakpoints, which act as boundaries between each piece. 
There have been a few studies on finding the number of breakpoints and their locations. 
% Rosen and Pardalos 
\cite{ rosen1986global} propose a method for finding the minimum number of equally spaced breakpoints within a given error tolerance, a sequential method is proposed in 
% Strikholm 
\cite{strikholm2006determining} for finding the number of breakpoints, and 
% Yang et al.\ 
\cite{yang2016mathematical} propose a discontinuous piecewise linear approximation and how to determine the optimal breakpoint locations.

Piecewise linear models have been used in different applications when modelling the structural shifts in data and forecasting based on the most recent behaviour in data is desired. We will discuss a PLR model, namely MARS, in more detail in Section \ref{plr}, where we discuss the supply and demand forecasting model used in this work. The MARS model has been used in healthcare for medical diagnosis using classification approaches \citep[see][]{butte2010validation, chou2004mining, yao2013novel, senthilkumar2013diabetes, lasheras2020methodology, serrano2020identification}. In the area of time series forecasting, 
% L{\'o}pez-Lozano et al.\
\cite{ lopez2019nonlinear} evaluate the generalizability of MARS for identifying thresholds for antibiotic consumption and 
% Katris
\cite{ katris2021time} studies MARS and other time series approaches for predicting the evolution of reported COVID-19 cases to track the outbreak in Greece.

\hfill \\ 
\textbf{Disaster Management and Resource Allocation}

In terms of resource allocation, the literature on emergency logistics involves disaster operations performed before or after the impact of a disaster where short-term post-disaster planning begins immediately. This includes activities among which relief distribution is more related to our scope of the study. Relief distribution consists of providing medical supplies, shelters, manpower, and other associated resources, and thus, resource allocation is a significant component \citep{caunhye2012optimization}. A number of resource allocation models for relief distribution that consider resource assignment without determining flow quantities among arcs focus on allocating equipment to amend maritime disasters, such as cleaning up oil spills. Unlike other disasters, oil spills create a single demand point, and studies in this area only consider direct equipment assignment to the spill areas \citep[see][]{psaraftis1986optimal, srinivasa1997procedure, wilhelm1997prescribing}.

\hfill \\ 
\textbf{Optimization Models and Resource Allocation}

Many studies have focused on developing allocation models for medical resources during infectious disease outbreaks. A dynamic linear programming model based on an epidemic diffusion model is introduced in
% Liu et al.\ 
\cite{liu2013dynamic} to allocate medical resources.
% Preciado et al.\ 
\cite{preciado2013optimal} analyze a networked version of a susceptible-infected-susceptible (SIS) epidemic model when different susceptibility levels are present. They propose a convex optimization approach for distributing vaccination resources in a cost-optimal manner and test their approach in a real social network. 
% Yarmand et al.\ 
\cite{yarmand2014optimal} consider two-phase vaccine allocation to different geographical locations. They capture each region’s epidemic dynamics for different vaccination phases by a two-stage stochastic linear program (2-SLP) model and show that their model helps to reduce vaccine production and administration costs.
% Yin and Büyüktahtakιn
\cite{yin2021multi} define two new equity metrics for fair resource allocation and develop a multi-stage stochastic programming vaccine allocation model and apply it to an Ebola virus disease case study. Furthermore, two resource allocation problems during outbreaks are discussed in 
% Preciado et al.\ 
\cite{preciado2014optimal}, where they use geometric programming to solve the problems. Following the work in \cite{preciado2014optimal}, 
% Han et al.\ 
\cite{han2015data} propose a data-driven robust optimization framework based on conic geometric programming. Their model is used to determine an optimal allocation of medical resources such as vaccines and antidotes and can help control an SIS viral spreading process in a directed contact network with unknown contact rates. 

\hfill \\ 
\textbf{Other Epidemiological Frameworks and Resource Allocation}

A number of works have investigated resource allocation frameworks using outbreak case studies. The problem of scheduling limited available resources between multiple infected areas is discussed in
% Rachaniotis et al.\ 
\cite{rachaniotis2012deterministic}, and their proposed deterministic scheduling model is studied in a case study of mass vaccination against A(H1N1)v influenza. A real-time synchronous heuristic algorithm is proposed in \cite{rachaniotis2017controlling} and is tested on the same case study as in \cite{rachaniotis2012deterministic}. 
% Sun et al.\ 
\cite{sun2014multi} focus on allocating patients and resources between hospitals located in a healthcare network and propose a multi-objective optimization model. They discuss the application of their model in an influenza outbreak case study. Finally, a large integer programming problem framework for optimally allocating a resource donation is introduced in
% Anparasan et al.\ 
\cite{anparasan2019resource}, and results of applying the framework to a 2010 cholera outbreak case study are reported. Closely related to our work is 
% Du et al.\ 
\cite{du2021data} where they study a multi-period location-specific resource allocation problem for cholera outbreak intervention. They consider a rolling time horizon and periodically determine an optimal intervention resource allocation strategy with their data-driven optimization approach. Also similar in spirit to our approach is 
% Bekker et al.\ 
\cite{bekker2023modeling}, who propose a model for making daily short-term predictions of the number of occupied ICU and clinical beds in the Netherlands due to COVID-19.

\hfill \\ 
\textbf{Motivation Behind Proposed Methodology}

With the aforementioned approaches in mind, we now discuss the motivation behind the prediction methodology that we consider for real-time short-term demand and supply forecasting for experimental products in clinical trials. We observed the following limitations with the application of the methodologies above to our problem setting:
\begin{legal}[leftmargin=0.3cm,itemindent=0.5cm,labelwidth=\itemindent,labelsep=0cm,listparindent= 0.5cm, align=left]
\item When an epidemic is first emerging, the epidemic state is only partially observable, and the parameters that the epidemiological models require are often indeterminate, as the disease information can only be obtained over a period of time and after sufficient cases are reported and required data is collected. Moreover, different geographical locations can show various characteristics in terms of the disease spread pattern. The estimates that compartmental models make are sensitive to the model's structure \citep{silal2016sensitivity}. Even in the presence of reasonable parameter estimates and simple model structures, real-time resource allocation requires the parameters to be periodically updated based on the disease spread rate and the number of people involved, whether susceptible or infected. Therefore, CHIME-like models may lead to poor approximation of the actual demand when used in a real-time setting where obtaining the most recent updated parameters may not always be possible. \\
\item The aforementioned demand forecasting models work best when large datasets are available and the models can capture the trend and seasonality, which is not possible during emerging epidemics. Furthermore, real-time demand forecasting can be challenging with any forecasting model if the demand is affected by external factors, such as population characteristics, geographical locations, operational procedures, guidelines, and governmental policies. As far as we are aware, none of the mentioned approaches are consistent with the challenges we introduced in Section \ref{Introduction}. It is quite difficult to directly apply these approaches to a real-time setting where short-term supply and demand forecasting is desired, there is a limited supply of resources, and the available data is sparse and shows fundamental changes in behaviour during different periods. This motivates us to avoid models that are reliant on a large number of parameters (as for CHIME-like models), as they require continual updates and they may not always yield accurate forecasts for resource supply and demand. Nonetheless, we seek a model that makes reasonable predictions even when facing such challenges. We will revisit these challenges in the case study in Section \ref{SimulationResultsplusInsights}. \\
\item More sophisticated demand forecasting models, such as ARIMA, can also be used for time series analysis, especially demand forecasting with univariate data. However, satisfaction of basic requirements for such models is necessary which might not be possible in our setting. For instance, ARIMA and ARMA work best in the presence of stationary data and the rule of thumb for these models suggests having at least $50$ but preferably more than $100$ observations \citep{box1975intervention}. We cannot meet that requirement, and as we start forecasting with almost no historical data, any forecasting model might fail to exhibit its best performance. We found a simple forecasting model such as MARS to be a good choice considering our non-stationary and limited data. \\
\item Our study is different from the aforementioned disaster management literature in the sense that although there is the common objective of resource allocation of limited resources, the time scale is different. Epidemics tend to be much longer than disasters such as earthquakes, and there is more uncertainty in the quantity of the effective resources due to the unknown nature of the disease. Furthermore, the evolution of epidemics might involve unexpected changes in supply and demand patterns, requiring suitable forecasting models to be incorporated into the resource allocation model.
\end{legal}

{This work focuses on \begin{enumerate*}
\item how the methodologies above are combined to solve a real-time short-term supply and demand forecasting and resource allocation problem, and \item how the combined methodology is applied to a clinical problem setting with various human interventions and limited historical data.\end{enumerate*}} We demonstrate that using MARS to determine inputs (supply and demand forecasts) to a resource allocation problem is an effective combination in an emerging epidemic setting. The prediction model in \cite{bekker2023modeling} consists of a linear programming model inspired by smoothing splines for predicting the arrivals and methods stemming from queueing theory to convert arrivals into occupancy. The motivation for choosing their model is similar to ours in the sense that it works with little historical data, which is a consequence of an emerging epidemic setting. To the best of our knowledge, no other study has tackled the problem of real-time multi-location allocation of scarce resources with limited historical data and without relying on epidemiological models.

\section{Data-driven Resource Allocation Model} \label{demandforecasting}

\subsection{Problem Description} \label{problemdescription}

Our work considers the patients of a COVID-19 case study (details in Section \ref{SimulationResultsplusInsights}) where the distribution of supply and demand is similar in terms of the population; however, there is a limited and highly-varied pattern for both in the collected data. The supply for this study comes from the CCP of the patients who have recovered from COVID-19, and donated blood. The demand comes from the patients who are currently infected with COVID-19 and require CCP, but will be considered as supply once they recover and donate CCP. 

\begin{figure*}[!htbp]
\centering
\includegraphics[width=1\textwidth]{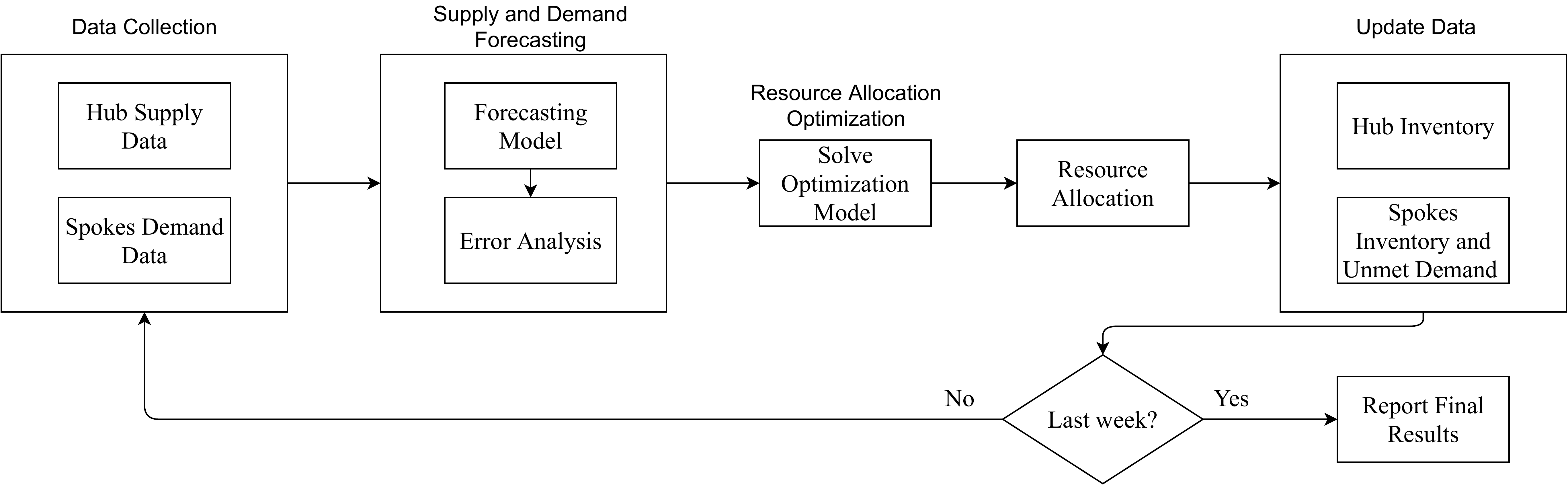}
\caption{Data-driven Resource Allocation Process}
\label{fig:flowchart}
\end{figure*}

Hub-and-spoke structures can help effective access to blood products in a public health network and clinical trials \citep{fernandez2022hub, roberts2018evaluation}. A hub-and-spoke structure was established for
our case study where CCP units were distributed from regional blood centers to hospital 
hubs for subsequent CCP allocation to patients. Therefore, our model also considers a hub-and-spoke structure where we have a centralized supplier (hub) that interacts with $H$ customers (spokes) and is responsible for satisfying their demand for $R$ types of resources. Figure \ref{fig:flowchart} shows a flowchart of our data-driven resource allocation process.

We are interested in a setting where limited resources must be allocated on a regular basis (e.g. every week) to the entities requesting them, the demand for the resources can be heterogeneous and arises from geographically dispersed locations. We forecast both supply and demand for the following reasons: \begin {enumerate*}  \item Common source: The same population defines both supply and demand, but at different points in time. \item Dependence between supply and demand: Changes in the behaviour of the disease naturally lead to correlated changes in supply and demand. \item Uncertainty in supply and demand: A classic resource allocation problem considers known resource quantities. However, we do not know the number of resources that are available to be allocated at every point in time and we face uncertainty in both supply and demand due to the unknown nature of the disease. The uncertainty in inventory collection also indicates the impact of CCP collection in our study, resulting in a resource collection and allocation model. \end {enumerate*}

We choose to work with the cumulative supply or demand as a cumulative sum of data samples over time is helpful in situations where one needs to smooth heterogeneous and sparse data and still make quantitative predictions for future supply and demand without altering the original data. 
% Ellaway 
\cite{ellaway1978cumulative} investigates the application of the cumulative sum technique in a neurophysiology study. The cumulative sum is shown to be a powerful technique for finding periods of change in data, as well as reducing real-time decision-making uncertainty. In our case, we forecast weekly supply and demand based on historical cumulative supply and demand data for a particular resource prior to resource allocations. In particular, we would like to fit a forecasting model weekly to our data where for each week, only the most recent week’s observation is added to the dataset, and we cannot modify the predictions the model previously made (as real-time allocations are made based on the forecasts). We then perform error analysis of the forecasts, and allocate the available resources to the customers in a fair manner by solving an appropriate optimization problem. We assume that both the supplier and customers can hold inventories of the resources and can use them to satisfy future demand. Once the supplier allocates resources to customers, the decision is final, and the resources cannot be reassigned. Thus, we need to update the inventories at the end of each week, and consider them when solving the optimization model in the next week.

We discuss a PLR supply and demand forecasting model and our resource allocation optimization model in more detail in Section \ref{supplyanddeamndforecasting} and Section \ref{optimization}, respectively.

\subsection{Supply and Demand Forecasting} \label{supplyanddeamndforecasting}

We would like to forecast the supply and demand for a particular week $t+1$ where we only have data available up to week $t$. Consider a dataset consisting of $t$ observations where $X$ is an array of elements $x_i$ ($i=1,2,{\dots},t$) representing the input variables (in our case, $X={1,2,3, {\dots},t}$). Consider $y$ to be a vector of observed (supply or demand) data samples $y_i$ ($i=1,{\dots},t$). We use Multivariate Adaptive Regression Splines (MARS), a PLR model, to forecast the supply and demand. 

\subsubsection{Multivariate Adaptive Regression Splines (MARS) } \label{plr}

Multivariate Adaptive Regression Spline (MARS) is a nonparametric regression approach that was introduced by 
% Friedman 
\cite{friedman1991multivariate}. The MARS model consists of a collection of simple linear models that can capture patterns and trends related to interactions and nonlinearities. MARS uses a series of piecewise linear pieces (splines) of different gradients. These pieces, also known as basis functions (BFs), are connected at positions called knots which allow thresholds, bends, and other departures from linear functions. A MARS model is specified as follows:

\[{\hat{y}}'=\beta _{0} +\sum _{p=1}^{P}\beta _{p}  \lambda _{p} (x),\] 

\noindent where $P$ is the number of BFs and each $\lambda _{p} (x)$ is a BF, which can be a spline function or the product of two or more spline functions. The parameters $\beta_0$ and $\beta_p$, $p=1,\ldots,P$ are estimated using the least-squares method. The basis functions are described as:

\[BF(x) =\{ \max (0,x-c_{p} ),\max (0,c_{p} -x)\}, \] 

\noindent where $c_{p} $ is the knot of the spline (threshold value).

A forward stage and a backward stage are considered in the MARS algorithm. In the forward stage, BF functions and their potential knots are chosen, which may result in a complicated and over parameterized model. In the backward stage, to prevent overfitting, the algorithm considers deleting the BFs in increasing order of the amount that they reduce the training error \citep{friedman1991multivariate}.

\subsubsection{Error Analysis and Model Enhancement} \label{erroranalysis}

Based on the slope of the last piece, the forecast supply and demand for week $t+1$ under MARS is simply:  

\[{\hat{y}_{t+1}}' ={\hat{y}_t}' +\frac{{\hat{y}_t}' -{\hat{y}_{t-1}}' }{x_{t} -x_{t-1} }.\]

We can improve the forecast value for week $t+1$ (${\hat{y}_{t+1}}'$) by calculating its forecast error \citep{jarrett1987study, lu2020improved}. We first fit an autoregressive (AR) model of order $l$ using Conditional Maximum Likelihood to the residual error (${\epsilon_i=y_i-{\hat{y}_i}'}$) data up to week $t$ and use it to forecast the error for week $t+1$:

\begin{equation}\label{eq:10}
\hat \varepsilon _{t+1} =a_{0} +a_{1} \varepsilon _{t} +a_{2} \varepsilon _{t-1} + {\cdots} +a_{l} \varepsilon _{t-l+1},  
\end{equation}

\noindent where ${a_0, a_1, {\dots} , a_l}$ are the coefficients obtained from the AR model and ${\hat{\epsilon}_{t+1}}$ is the forecast error for week $t+1$. We finally calculate $ {\hat y_{t+1} = {\hat{y}_{t+1}}' + \hat \varepsilon _{t+1}}$, i.e., the improved supply or demand forecast value for week $t+1$, and use it as our final forecast value for that week.

\subsubsection{Challenges} \label{challenges}

We now discuss general challenges that may be faced when MARS is employed in an online manner: 

\begin{itemize}  
 
\item Predictions made using MARS may degrade if there is a sudden transitory change in data. For instance, we found that holidays may affect the amount of data collected in a particular week but the data follows its previous pattern after the holidays have passed. We discuss this issue in greater detail in the case study discussed in Section \ref{SimulationResultsplusInsights}. 
\item MARS is fitting piecewise linear models, and thus the slope of the segment that ends with the most recent observation has a significant effect on the forecast for the next week (it may lead to a large underestimation/overestimation). \end{itemize} 

\begin{figure*}[!htbp]
\centering
\subfloat[cumulative forecast demand for week $21$]{%
\resizebox*{.49\textwidth}{!}{\includegraphics{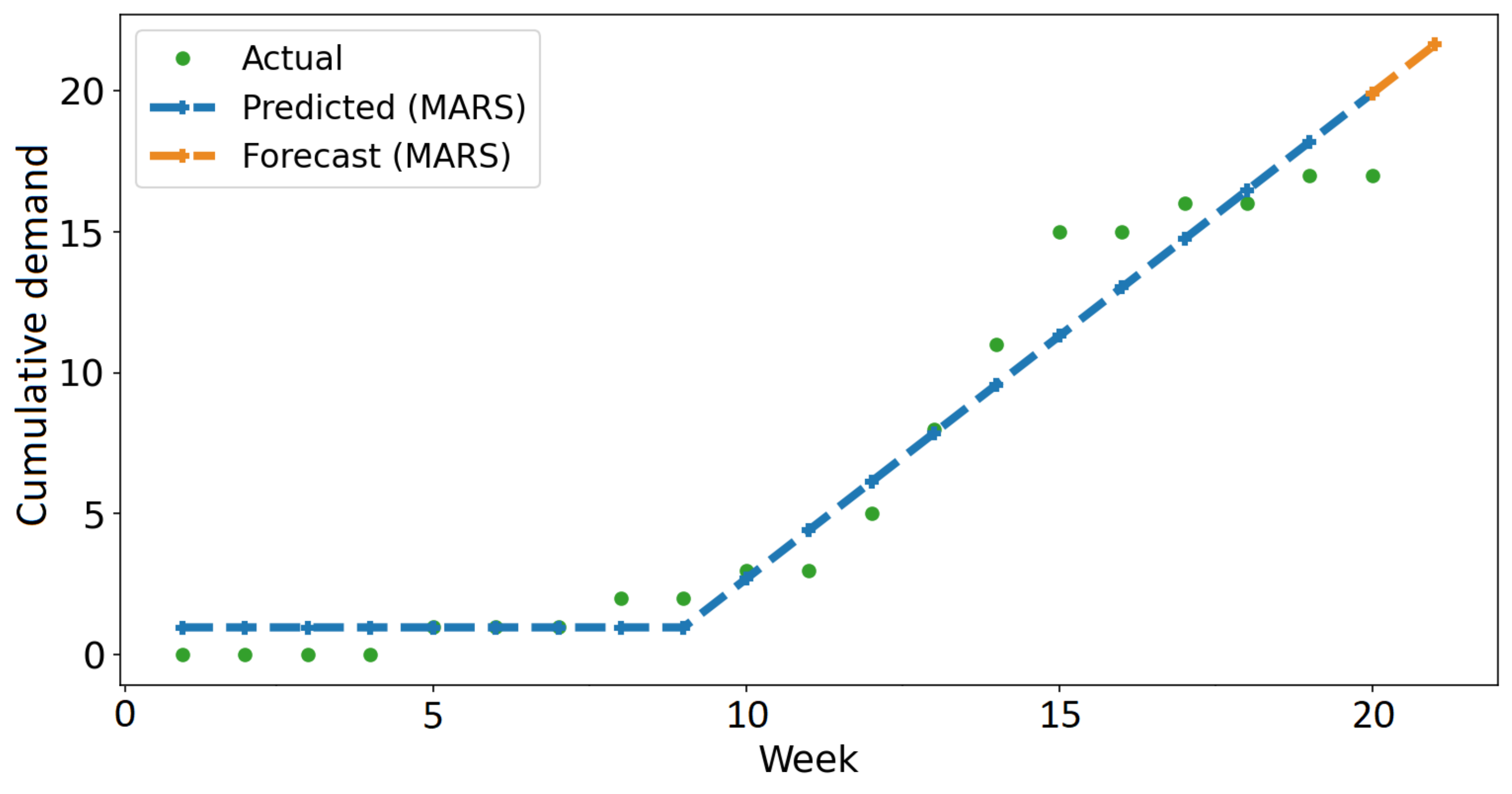}}\label{fig:7}}\hspace{5pt}
\subfloat[cumulative forecast demand for week $22$]{%
\resizebox*{.49\textwidth}{!}{\includegraphics{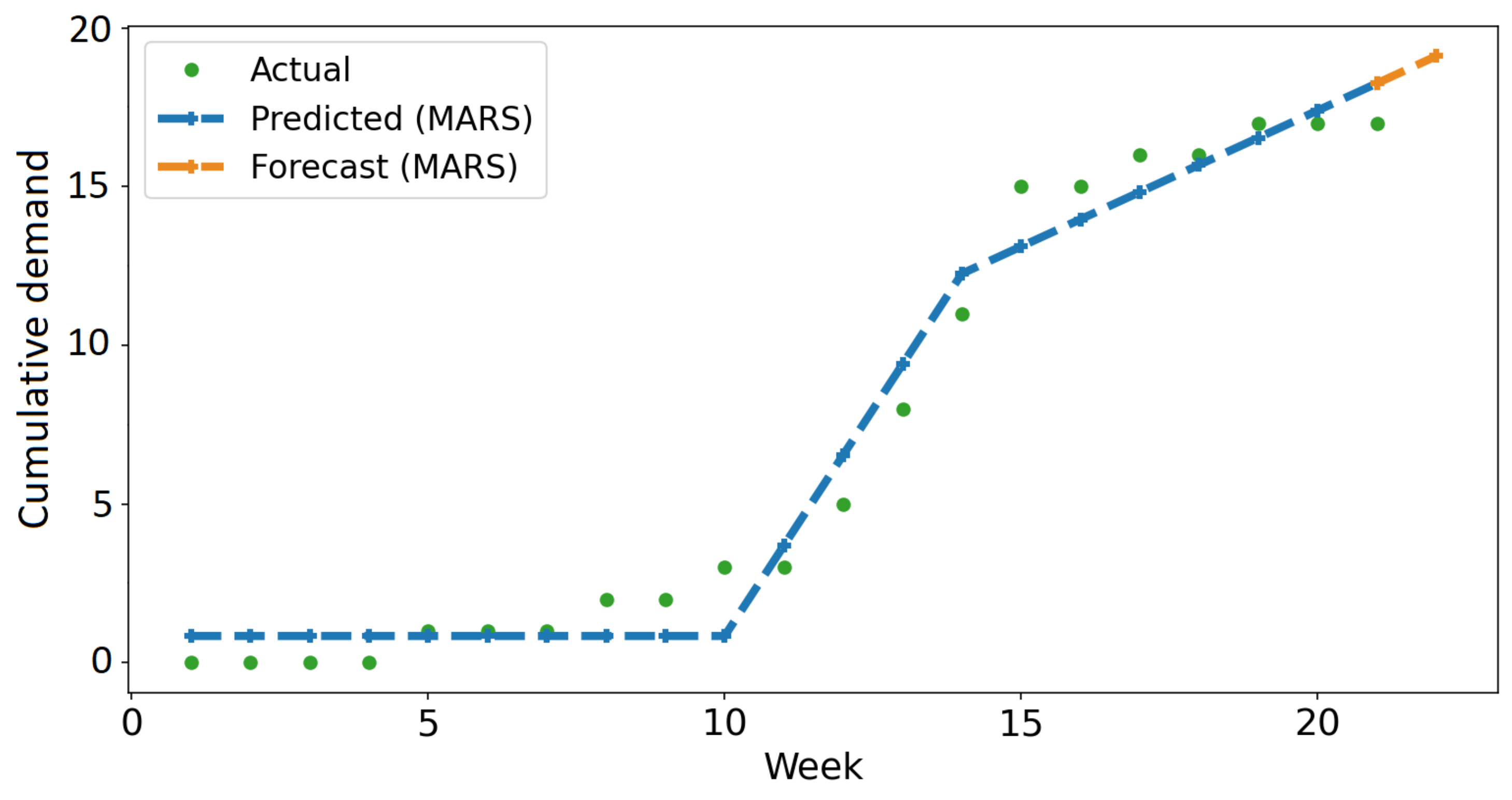}}\label{fig:8}}
\caption{Forecasting a negative non-cumulative value} \label{fig:neg} 
\end{figure*}

Although working with the cumulative sums for forecasting future supply and demand can help address the issue of data sparsity, a particular challenge arises when cumulative sums are employed in an online setting. Consider a situation where the demand before week $t$ has a steep slope resulting in a (relatively) high forecast for week $t$. However, upon observing the demand for week $t$, one finds out that the actual demand was considerably lower. Thus, the slope that affects week $t+1$'s forecast demand will be less steep than what it was when only data up to week $t-1$ was available. This may result in the forecast cumulative demand for week $t+1$ being lower than the previously forecast cumulative demand for week $t$, which is not possible. For instance, Figure \ref{fig:7} and Figure \ref{fig:8} demonstrate the cumulative forecast demand obtained by MARS for week $21$ and week $22$, respectively. The cumulative forecast demand for week $21$ and week $22$ are $22$ and $19$, respectively, under MARS, which cannot happen in practice. This is one consequence of considering an online setting where modifying the predictions is not possible as real-time decisions are made. The sparser the data, the more this issue can affect the model's predictions. A possible solution for dealing with such situations is to assume that the cumulative forecast value for week $t+1$ is equal to the cumulative forecast value for week $t$. We will observe this issue in our case study discussed in Section \ref{SimulationResultsplusInsights} and deal with it in the suggested manner.

% \begin{figure*}[!htbp]
% \begin{subfigure}{.5\textwidth}
% \centering
% \includegraphics[width=1\textwidth]{fig2a.pdf}\caption{cumulative forecast demand for week $21$}\label{fig:7}
% \end{subfigure}
% \begin{subfigure}{.5\textwidth}
% \centering
% \includegraphics[width=1\textwidth]{fig2b.pdf} \caption{cumulative forecast demand for week $22$}\label{fig:8}
% \end{subfigure}
% \caption{Forecasting a negative non-cumulative value} \label{fig:neg} 
% \end{figure*}

\subsection{Resource Allocation Optimization} \label{optimization}

We allow for demand for resource $r$ to be satisfied by resource $r'$. This can be represented with a matrix $C$ of size $R \times R$, where an element ($r$,$r'$) of $C$ is $1$ if demand for resource $r$ can be satisfied by resource $r'$ and is $0$ otherwise. Furthermore, the supply used in our MIP model for resource $r$ for week $t$ is constrained as follows:

\begin{equation}\label{eq:11}
  c_{r,t}=\left\{
  \begin{array}{@{}ll@{}}
    \hat s_{r,t}, & \text{if}\ \hat s_{r,t} \le s_{r,t} \\
    s_{r,t}, & \text{otherwise},
  \end{array}\right.
\end{equation} 

\noindent where $s_{r,t}$ and $\hat s_{r,t}$ are the actual and forecast supply for resource $r$ on week $t$, respectively. The second case in \eqref{eq:11} indicates that we cannot allocate resources beyond the actual available supply. We will formulate our resource allocation as a Mixed Integer Program (MIP). To do so, we require the following notation:

\textbf{Indices} 

\begin{conditions*}
 t     &  index of time periods, $t=1,{\dots},T $ \\
 h     &  index of customers, $h=1,{\dots},H$ \\   
 r,r' &  index of resource, $r,r'=1,{\dots},R$
\end{conditions*}

\textbf{Data} 
\begin{conditions*}
 c_{r,t}     &  the amount of resource $r$ available at the supplier for assignment at time $t$ (see \eqref{eq:11} above) \\
 i_{r,h,t-1}     &  the inventory of resource $r$ stored at customer $h$ at time $t-1$  \\   
 \hat d_{r,h,t} &  the estimated demand for resource $r$  by customer $h$ at time $t$
\end{conditions*}

\textbf{Decision variables} 
\begin{conditions*}
  
 v_{r,r',h,t} &  the number of units of resource $r'$ assigned to customer  $h$ to satisfy demand for resource $r$  at time $t$. We only consider the set of $v_{r,r',h,t}$ that correspond to $C(r,r')=1$.
 
\end{conditions*}

We formulate our resource allocation problem as follows:
         
\textbf{Objective function} 
\begin {equation}\label{eq:1}
\min {\mathop{\max }\limits_{h:\sum_{r=1}^R \hat{d}_{r,h,t} > 0}} \frac{\sum _{r=1}^{R} (\hat{d}_{r,h,t} -\sum _{r'=1}^{R}v_{r,r',h,t} -i_{r,h,t-1})  }{\sum _{r=1}^{R} \hat{d}_{r,h,t} }.
\end {equation}

\textbf{Constraints}                          
\begin {equation}\label{eq:2}
\sum _{h=1}^{H}\sum _{r=1}^{R}v_{r,r',h,t} \le c_{r',t}   {\rm \; ,\; } \forall {\rm \emph r'},
\end {equation}
\begin {equation}\label{eq:3}
v_{r,r',h,t} \ge 0{\rm \; and\; integer\; valued}, 
\end {equation}
\begin {equation}\label{eq:4}
\sum _{r'=1}^{R}v_{r,r',h,t}  \le \hat d_{r,h,t} {\rm \; \; \; ,\; }\forall {\rm \emph r\; ,\; }\forall {\rm \emph h}.
\end {equation}
Constraint \eqref{eq:2} prevents the over-allocation of available resources. Constraint \eqref{eq:3} ensures the integrality and non-negativity of the resource allocations and constraint \eqref{eq:4} keeps the allocation to each customer below the corresponding estimated demand. If all of the estimated demand at time $t$ can be met, any excess supply is held in inventory at the hub.

The objective function \eqref{eq:1} captures our notion of fairness: minimizing the largest ratio of unmet demand over all customers. It could be modified to capture other notions of fairness. For example, one can penalize larger deviations between the demand and what is allocated by using a quadratic objective function, which would create a Mixed Integer Quadratic Program (MIQP). If the problem size becomes too large, one could also look at Linear Programming (LP) relaxations of the problem.
\section{The CONCOR-1 trial: A case study for a proposed application of the resource allocation model}\label{SimulationResultsplusInsights}

The Randomized, Open-Label Trial of CONvalescent Plasma for Hospitalized Adults With Acute COVID-19 Respiratory Illness (CONCOR-1) was an RCT involving 72 academic and community sites across Canada, the USA, and Brazil \citep{begin2021convalescent}. The randomization in this RCT was performed at a ratio of 2:1 allocation to receive CCP or standard of care for a planned study population of $1200$ patients, stratified by age ($<60$ and  $\ge 60$ years). The first CCP unit for the trial was collected on April $24$, $2020$, and the first patient was randomized on May $14$, $2020$. The trial ceased on January $29$, $2021$ with a total of $940$ randomized patients. The objective of the trial was to assess whether transfusing CCP reduces deaths or the proportion of patients requiring intubation at day 30 compared to standard of care for hospitalized COVID-19 infected adult patients \citep{begin2021convalescent}. Canadian Blood Services (CBS) is responsible for collecting the CCP units and is the national blood supplier across all provinces in Canada except Qu\'ebec. The supply and demand management of the CCP products for patients enrolled in the trial was conducted by CBS and the CONCOR-1 team at the McMaster Centre for Transfusion Research. In what follows, as we are evaluating our approach, we will take the viewpoint that the trial is in progress.

Canada's blood supply chain network is currently centralized and comprises two levels: regional CBS distribution sites and hospital blood banks. Nine CBS blood distribution sites are currently located across Canada, and each centre attempts to meet the CCP demand from the hospital blood banks in its network. We use all available data from the trial which comes from $18$ hospital hubs, $30$ hospital sites, and $8$ CBS distribution sites.

CCP is stored frozen and ideally must be transfused as soon as possible after being thawed, but at most within five days \citep{bloch2021guidance}. A patient who is randomized to the CCP arm in the trial requires a single dose of approximately $500\ml$ or two doses of $250\ml$ (from a single or two different donors) and the CCP unit is transfused to the patient within the first 24 hours after randomization \citep{begin2021convalescent}.

It is challenging for CBS to make decisions on CCP allocation, because \begin {enumerate*}  \item the CCP supply is limited and restricted by manufacturing policies and may not meet the total CCP demand, \item the trial involves hospitals from geographically dispersed locations and multiple blood distribution centres, \item the demand for CCP exhibits heterogeneity between different geographic regions, \item there is limited knowledge or historical data about the disease demographics making it difficult to forecast the supply and demand of CCP products, \item there are specific clinical requirements for administrating CCP transfusions, such as specific product dose, ABO blood group compatibility, and medical condition requirements, and \item the decisions must be made in real-time and once the CCP units are shipped to a hospital hub, redistribution to other hospital hubs is undesirable. Hence, the decision for every unit matters. The key observations of our work suggest that our proposed data-driven MIP model can help balance the supply and demand of CCP products and lead to a fair allocation of limited CCP products among hospital hubs. \end {enumerate*} 

\begin{figure*}[!htbp]
\centering
\includegraphics[width=0.75\textwidth]{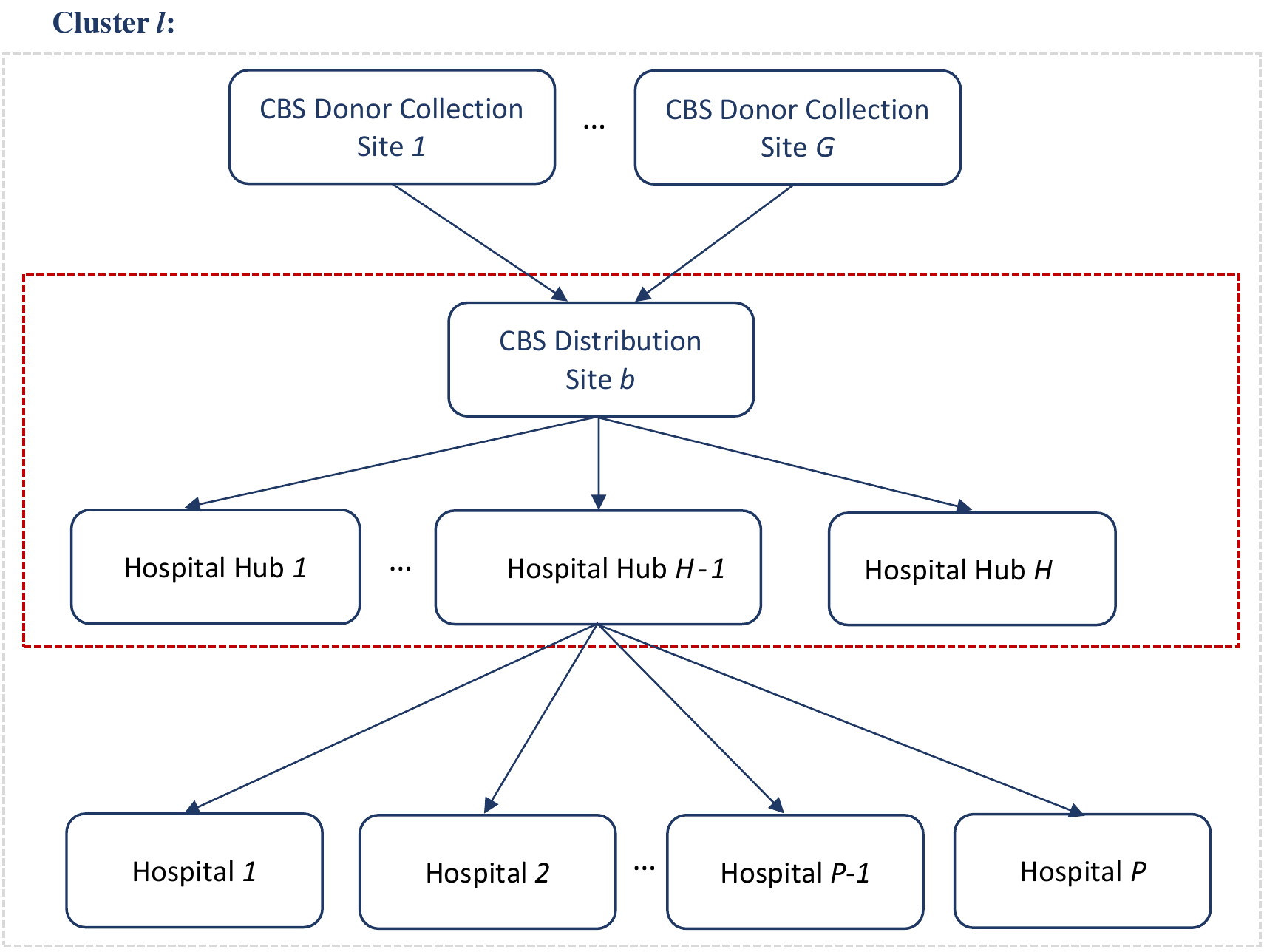}
\caption{CCP allocation network}
\label{fig:network}
\end{figure*}

The underlying network of our case study is shown in Figure \ref{fig:network}. The corresponding definitions and notations are listed as follows:

\begin{itemize}
\item $l$: represents a cluster and is defined as a geographic region with only one CBS distribution site but one or more CBS donor collection sites, hospital hubs and individual hospital sites.
\item $g$: represents an individual CBS donor collection site in cluster $l$, $g=1,{\dots},G$.
\item $b$: represents the CBS distribution site in cluster $l$.
\item $h$: represents an individual hospital hub under CBS distribution site $b$ in cluster $l$, $h=1,{\dots},H$.
\item $p$: represents an individual hospital site for hospital hub $h$ in cluster $l$, $p=1,{\dots},P$.
\end{itemize}

The structure in Figure \ref{fig:network} is a hub-and-spoke structure (a CBS distribution site is a hub and its underlying hospital hubs are the spokes) and is consistent with our resource allocation MIP model proposed in Section \ref{optimization}. CBS distribution sites can centrally decide to reallocate the blood products if there is excess supply in a particular CBS region. Thus, we only consider a single CBS distribution site in our optimization problem. Based on the agreements between CBS and Héma-Québec, the blood supplier in Qu\'ebec, less common blood groups (blood groups AB and B) can be shared \citep{begin2021convalescent}.

The CBS distribution site provides CCP to hospital hubs that in turn allocate units to other hospital sites in the area. We only focus on the allocations from the CBS distribution site to the hospital hubs.

As of May $2021$, the distribution of ABO blood groups in Canada is O:$46\%$, A:$42\%$, B:$9\%$, and AB:$3\%$ \citep{whatscbs}. In general, plasma for AB and O blood groups are universal donor and recipient, respectively. In practice, transfusing blood-specific units is prioritized. In the CONCOR-1 trial, due to the limited resources for B and AB plasma, patients with O and B blood groups can receive A and AB plasma, respectively, when an exact match is not available.

The dataset that we work with includes the CBS distribution site's available CCP units after shipment at an aggregate level for each blood group on specific dates starting from September $1$, $2020$, up to January $25$, $2021$. It also contains data for the received CCP units from the CBS distribution site for each hospital hub, whether randomized to a patient or stored in inventory, from May $11$, $2020$, up to January $25$, $2021$. Furthermore, CCP-related information such as product dose, ABO group, and whether a unit was broken or leaking after being thawed for transfusion at a hospital hub are recorded. Table \ref{tab:1} provides a summary of the dataset's attributes, their description, and their format.

\begin{table*}[htbp]
\captionsetup{font=large}
\caption{Dataset description}
\centering
\small
\begin{tabular}{p{1cm}p{2.5cm}p{8.2cm}p{1.00cm}}
\hline 
Dataset & Attribute & Description & Format \\ \hline
\multirow{7}{*}[+1ex]{\rotatebox[origin=c]{90}{CBS supply data}} 
& date & CBS distribution site's aggregate available units after shipment on a date & Date \\  
 & A & Total number of blood group A CCP ($250\ml$) units & Integer \\ 
 & AB & Total number of blood group AB CCP ($250\ml$) units & Integer \\ 
 & B & Total number of blood group B CCP ($250\ml$) units & Integer \\  
 & O & Total number of blood group O CCP ($250\ml$) units & Integer \\ 
 & Total & Total number of CCP ($250\ml$) units & Integer \\ \hline
\multirow{8}{*}[-6ex]{\rotatebox[origin=c]{90}{Hospital Hub data}} & hospitalhub\_ID & Unidentifiable unique ID of a hospital hub & String \\  
 & receiveddate & Hospital hubs received CCP units from the CBS distribution site on a date & Date \\  
 & DNL & The CCP unit's de-identified ID & String \\ 
 & productABOgroup & The ABO blood group of the CCP unit & String \\ 
 & productdose & The CCP unit dose with $1$ indicating $250\ml$ units and $2$ indicating $500\ml$ units & Integer \\ 
 & matched & Boolean variable with 1 indicating the CCP unit was matched with a randomized patient (indicating a demand), $0$ if the unit was stored in the hospital hub's inventory& Boolean \\ 
 & thawed & Boolean variable with $1$ indicating that the unit broke when thawing, $0$ otherwise & Boolean \\ \hline
\end{tabular}
\label{tab:1}
\end{table*}

We first create a cumulative weekly dataset of the number of new $500\ml$ units assigned to randomized patients at each hospital hub (considered as their CCP demand) and the total received units from CBS for each resource. We cumulatively sum hospital hubs' weekly received CCP units and CBS weekly inventory difference to calculate CBS cumulative new weekly supply. To maintain and promote efficient CCP allocation, prior estimation of the hospital hub's demand and CBS supply is necessary. One could either translate the CHIME model outputs to CCP supply or demand, or build a CHIME-like model that incorporates our variables. However, we found it more effective to work with the supply and demand directly and use PLR forecasting models. One reason is that the proportion of patients consenting to the trial also affects the total CCP demand. The consent rate can be affected by many factors such as a patient's religion and other competing treatment strategies. So, the patient population may not be the same as the population considered in the CHIME or CHIME-like models. Furthermore, given the limited and highly-varied pattern of CCP supply and demand in our dataset, PLR models appear to be an effective tool for forecasting CCP units.

\subsection{Model Assumptions} \label{simulationmodelassumptions}

We have created an environment for weekly allocation of CCP units from the CBS distribution site to the hospital hubs participating in the CONCOR-1 trial. Our dataset contains CBS supply data from a later date than when the first patient was randomized to receive a CCP unit. So, we consider the week that CCP supply was first reported as the starting week in our model (week of August $31$, $2020$). The last week in the dataset is the week of January $25$, $2021$, so the duration $T$ of our model is $22$ weeks. We combined hospital hubs that are very close in distance, under the same distribution network and with few COVID-19 patients. Small hospitals in distant areas with a very low number of hospitalized COVID-19 patients were removed from the study dataset.

Table \ref{tab:2} shows the actual supply and demand ($500\ml$ units) reported for each resource and in total in this period. There are a total of $15$ A, $14$ O, $9.5$ B, and $14$ AB $500\ml$ CCP units received from Héma-Québec recorded in the dataset. The numbers in Table \ref{tab:2} are only associated with CBS supply  (excluding any units received from Héma-Québec). We observe in the data that $4.5$ A, $3$ O, $2$ B, and $0.5$ AB $500\ml$ CCP units are unused at the hospital hubs because they were leaking or broken after being thawed for transfusion. These units are not included in our model as the wastage due to the thawed CCP units is negligible.

\begin{table}[htbp]
\centering
\begin{tabular}{llllll}
\hline
CCP Units ($500\ml$) & A& O & B & AB & Total \\ \hline
Supply & $157$ & $84.5$ & $29$ & $26$ & $296.5$  \\
Demand & $131.5$ & $94.5$ & $34.5$ & $33.5$ & $294$ \\ \hline
\end{tabular}
\caption{CBS CCP supply and CCP demand (August $31$, $2020$ - January $25$, $2021$)}
\label{tab:2}
\end{table} 

We assume that we have no information on the actual CCP supply and demand when deciding on how to allocate the CCP units. Thus, we perform weekly forecasts considering only the available data up to that week. The open source \emph {scikit-learn-contrib} library
\emph {py-earth} \citep{rudy2016py} is used for applying the MARS model, where the maximum degree of interaction terms generated by the forward stage (the $max\_degree$ parameter) is set to $2$ to better deal with the nonlinearities in the data, and the remaining parameters follow the default values. We are not performing any validations because our dataset is sparse and the performance of MARS cannot be significantly enhanced as using the cumulative sums has already smoothed the dataset. In general, a $k$-fold cross validation can be used with MARS to obtain a less biased estimate. We start making predictions four weeks from the start of available data to decrease the likelihood of severe underestimation or overestimation. Finally, a lag $l=1$ is used in the autoregressive model of residual errors in \eqref{eq:10}. This choice is due to a week's supply and demand tending to be closer to the amounts for the most recent week, as well as trying to avoid sudden changes that we might face when considering a larger lag.

We deal with the challenges that might occur in our forecasting process, as discussed in Section \ref{challenges}; if a negative non-cumulative forecast value on week $t+1$ arises, we instead set the cumulative forecast value equal to the cumulative forecast value for week $t$. Furthermore, we account for expected sudden transitory changes in data, such as weeks containing holidays. We observe that the hospital hubs' requested CCP units are relatively lower on weeks containing the Christmas, Boxing Day, and New Year's Day statutory holidays. This issue is due to fewer working days or reduced staffing for these weeks. Thus, it is important to account for these weeks as they can be falsely detected as breakpoints. We have chosen the value of $0.7$ as a reasonable adjustment factor and multiply the slope of forecast line going through the weeks containing Christmas, Boxing Day, and New Year's Day holidays by this factor. One might need different adjustment factors depending on the extent the supply and demand are affected on a particular holiday.

From this point on, wherever we refer to MARS, we are considering the model's forecasts after accounting for the mentioned challenges and performing error analysis. Because CCP is stored in $250\ml$ units, we allocate $250\ml$ units from CBS to the hospital hubs in our model. The forecast values are decimal values; therefore, we round the supply and demand forecast values to the nearest integer so that all values correspond to $250\ml$ units. All the results reported in the figures and tables correspond to $500\ml$ units.

We are interested in making a fair allocation of different ABO blood group CCP units among the hospital hubs while minimizing their unmet CCP demand proportions. We consider four resources ($R=4$) of A, O, B, and AB and examine two compatibility matrices $C$ for assigning the CCP units in our MIP problem , as described in Section \ref{optimization}: \begin {enumerate*} \item the identity compatibility matrix, and \item the ABO compatibility matrix used in the CONCOR-1 trial which allows the transfusion of A and AB plasma to patients with O and B blood groups, respectively, when the same blood group is not available (CONCOR-1 compatibility matrix).\end {enumerate*} For our primary study, we use the forecast supply and demand for each of the resources in our model for each week. Furthermore, to evaluate different allocation scenarios, we also forecast the aggregate supply and demand for each week and consider the distribution of ABO blood groups in Canada as the probabilities for calculating the forecast CCP for each resource \citep{zietz2020associations}. For this sensitivity analysis, we use the following probabilities based on the distribution of Canadian blood groups on each run of our model for generating the supply and demand for each blood group: $w_{c_{A}}=w_{h_{A}}=0.42$, $w_{c_{O}}=w_{h_{O}}=0.46$, $w_{c_{B}}=w_{h_{B}}=0.09$, and $w_{c_{AB}}=w_{h_{AB}}=0.03$, where $w_{c_r}$ and $w_{h_r}$ are the supply and demand probabilities for blood group $r$, respectively. The goal of the sensitivity analysis is to determine the performance of the model in situations where it is only possible to forecast the aggregate amount of resources in terms of supply and demand. This analysis also helps gain insights as to the generalizability of the results of our primary study.

We assume that both CBS and hospital hubs can store the excess units at the end of each week to use them in later weeks. We solve our MIP model based on forecast supply and demand for each resource, and the actual inventories held at CBS and hospital hubs on the week under study. We assume that the hospital hubs can use a compatible CCP unit according to the CONCOR-1 compatibility matrix (if available) for a patient when the same blood group CCP is not available. We examine both identity and CONCOR-1 compatibility matrices for solving our MIP model; however, at the inventory level of the hospital hubs, only the CONCOR-1 compatibility matrix is used. We shall see later that this is a good combination for the situations where forecast supply and demand are used, and the forecasting errors are not too large. The identity compatibility matrix at the MIP level prevents issues in terms of greedy allocation of scarce resources, and the CONCOR-1 compatibility matrix at the inventory level of the hospital hubs allows for the efficient compensation of forecasting errors. 

Because we know the actual demand for a particular week only after that week has passed and because the supply is limited, we might not fully meet all demands. In these cases, the unmet demand is carried over to the next week. We note that the optimization model ensures that the CCP units of a resource allocated to a hospital hub are never more than its forecast demand. Excess units stored in hospital hubs' inventories are due to the possible difference between the actual and forecast values.

A total of $m$ runs are used to calculate the mean final unmet demand ($\overline{u}_{T} $) and the mean ratio of final unmet demand to total demand ($\overline{z}_{T} $) (for ease of reading, we will subsequently refer to these quantities as unmet demand and unmet demand ratio, respectively) for each hospital hub: 
\[\overline{u}_{T} =\frac{\sum _{r=1}^{R}u_{T_{r} }  }{m} \] 
where the duration of our model is $T=22$ weeks, and $u_{T_{r} } $ is the hospital hub's final unmet demand for resource $r$ and,
\[\overline{z}_{T} =\frac{\sum _{r=1}^{R}u_{T_{r} }  }{m\sum _{t=1}^{T}\sum _{r=1}^{R}d_{t_{r} }   } \] 
where $d_{t_{r} } $ is the hospital hub's actual newly-added demand on week $t$ for resource $r$.

\subsection{Results} \label{simulationmodelassumptions}

\begin{figure*}[htbp]
\centering
\subfloat[Total units]{%
\resizebox*{.98\textwidth}{!}{\includegraphics{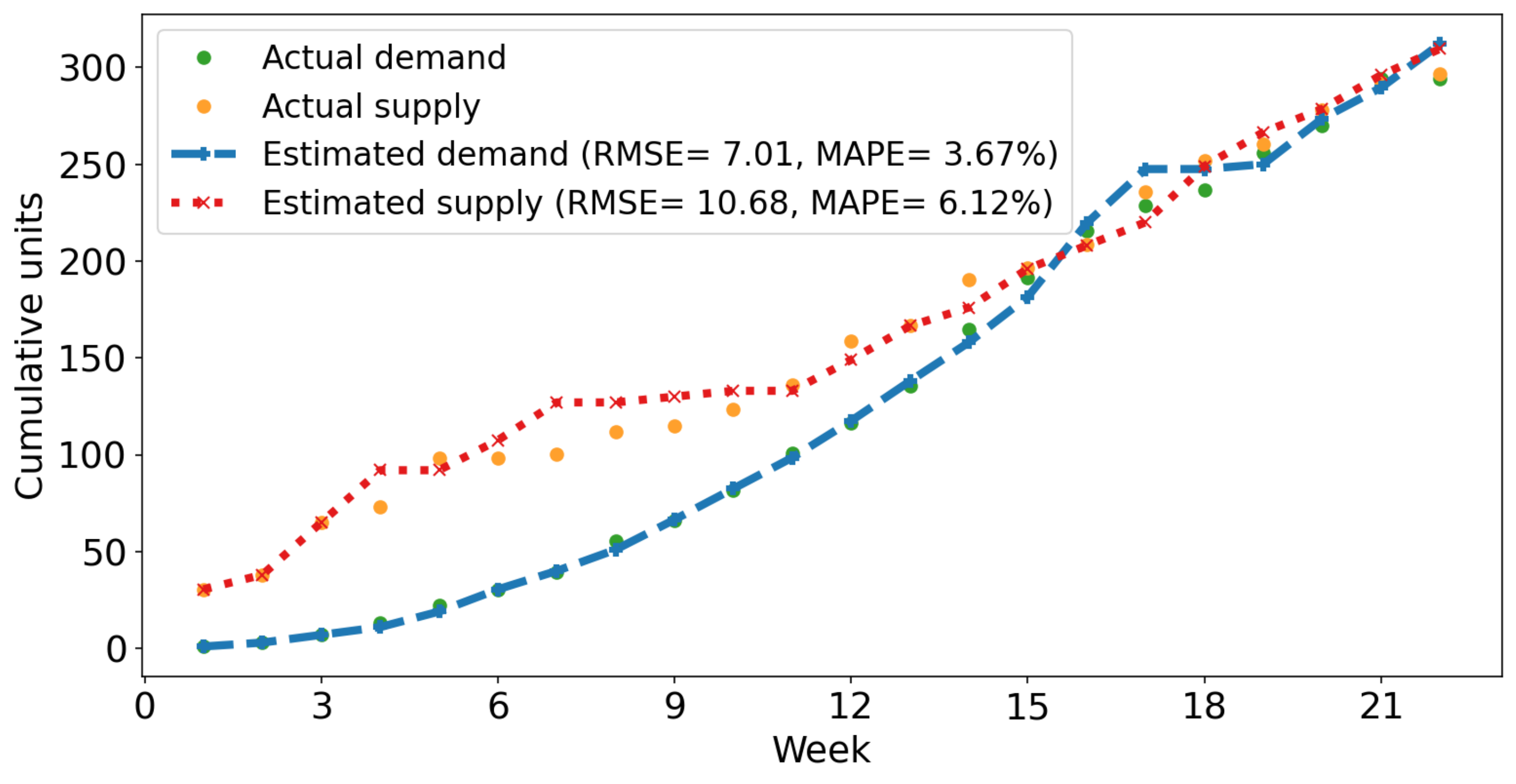}}\label{fig:totalsupply}}\hspace{5pt}
\centering
\subfloat[A units]{%
\resizebox*{.49\textwidth}{!}{\includegraphics{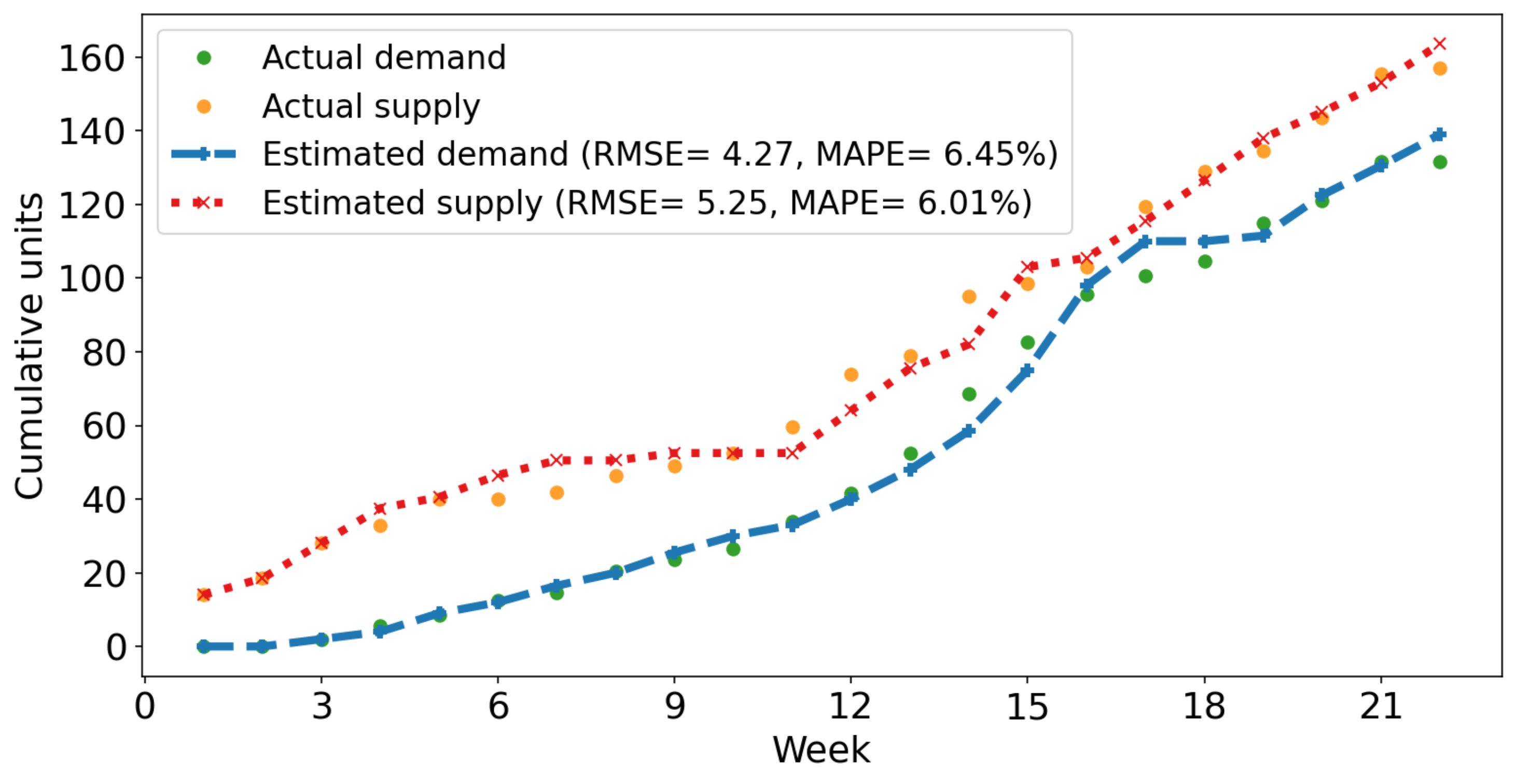}}\label{fig:asupply}}
\centering
\subfloat[O units]{%
\resizebox*{.49\textwidth}{!}{\includegraphics{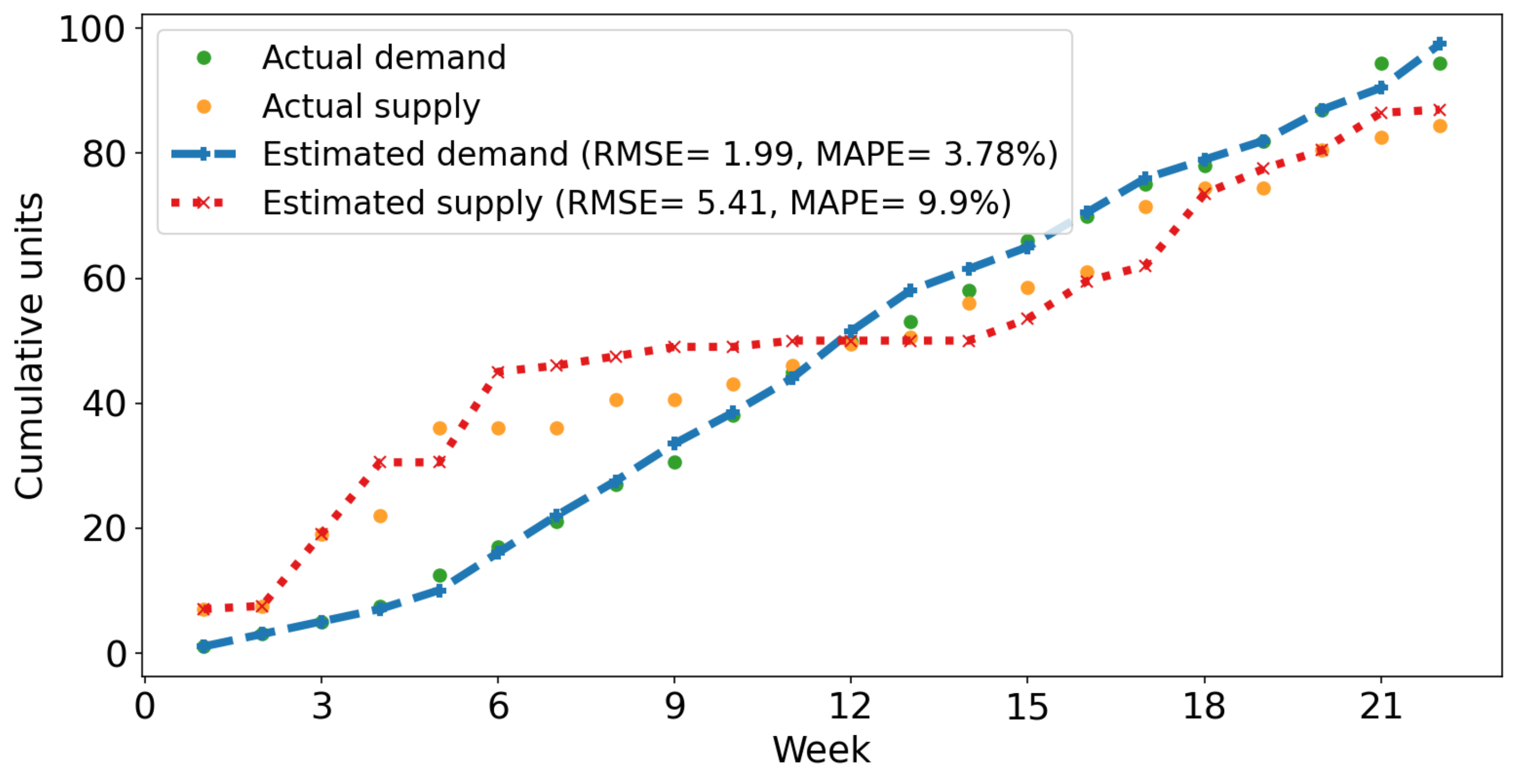}}\label{fig:osupply}}\hspace{5pt}
\centering
\subfloat[B units]{%
\resizebox*{.49\textwidth}{!}{\includegraphics{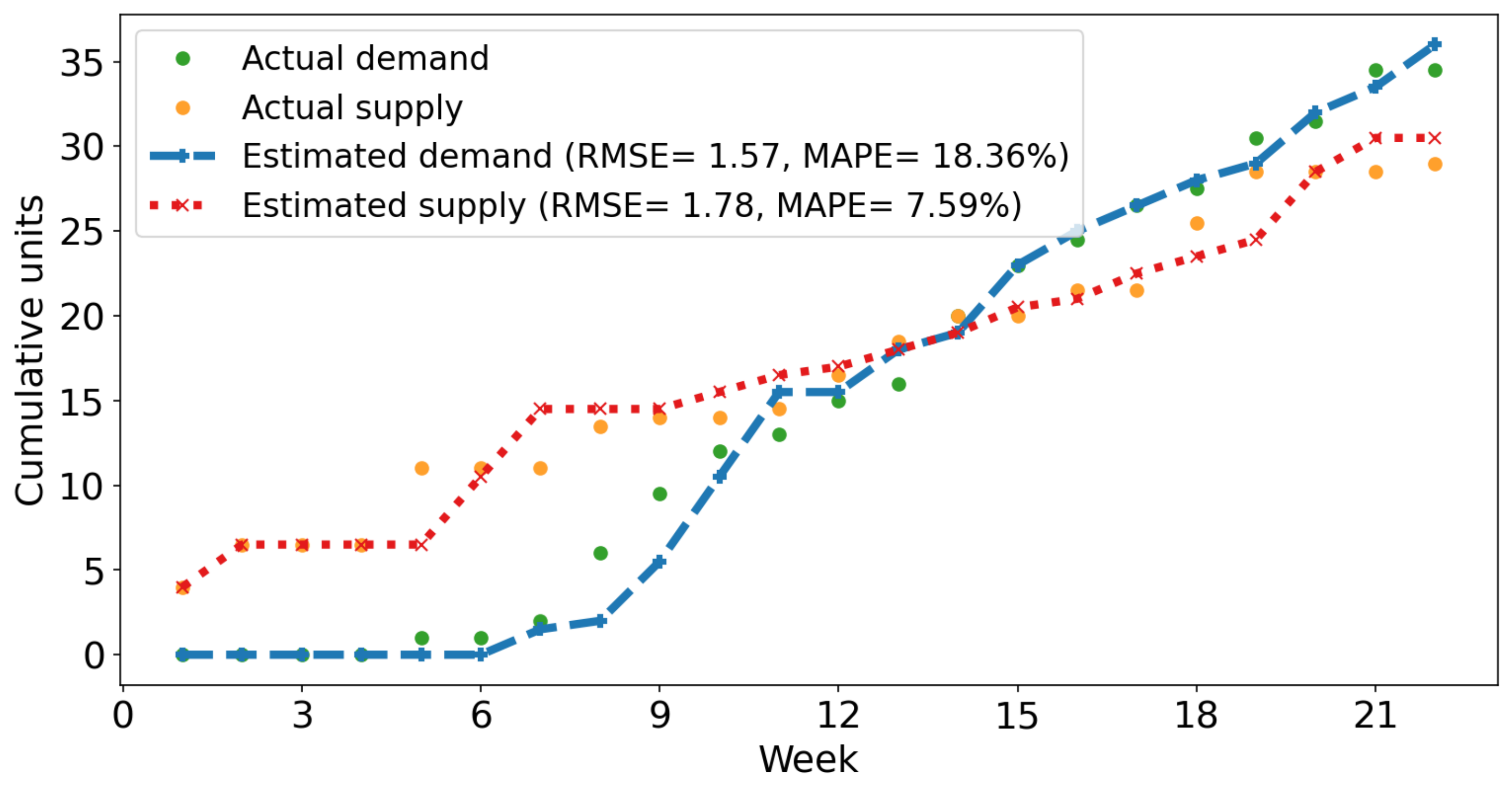}}\label{fig:bsupply}}
\centering
\subfloat[AB units]{%
\resizebox*{.49\textwidth}{!}{\includegraphics{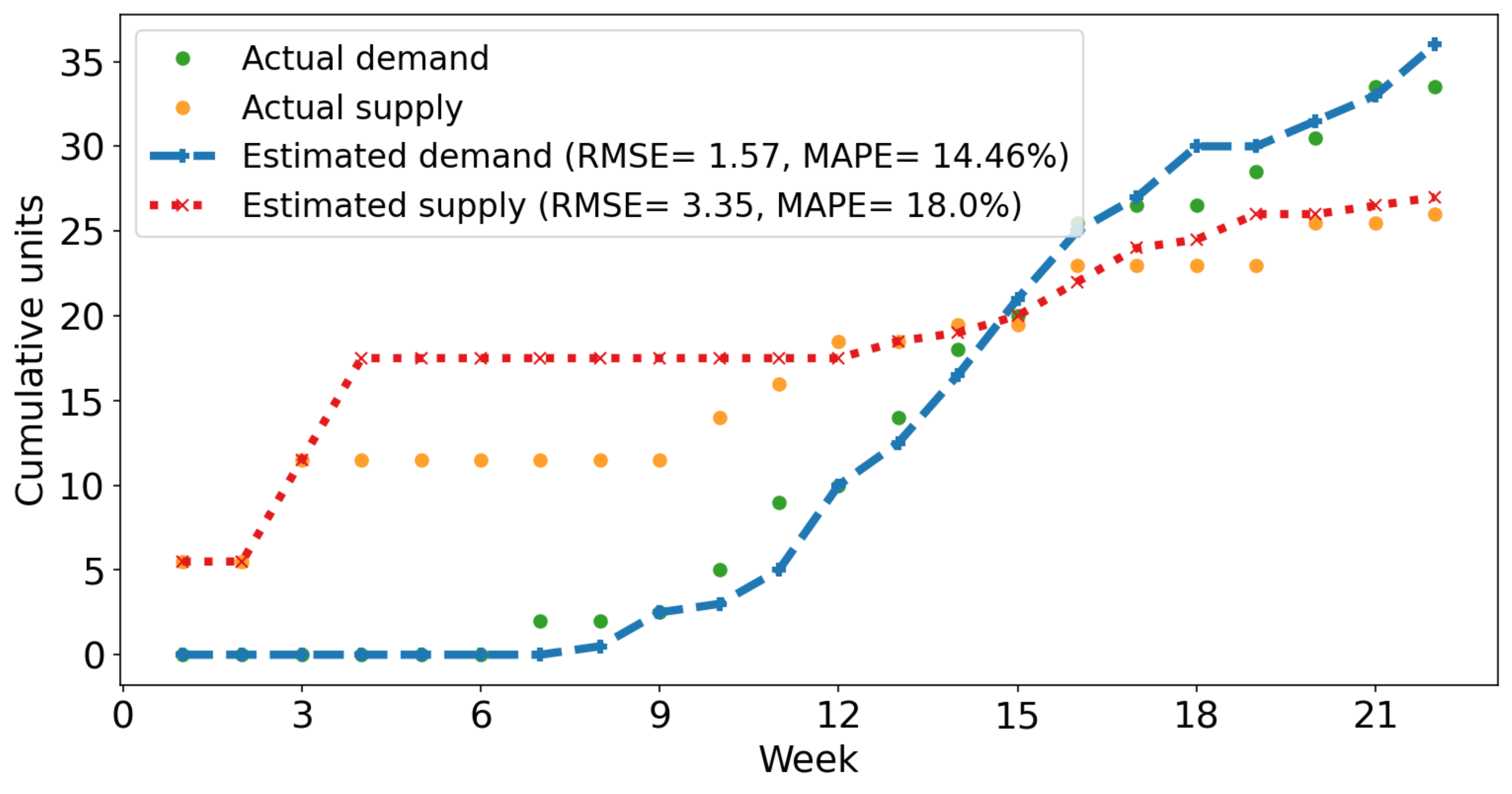}}\label{fig:absupply}}
\caption{Model performance in forecasting CCP supply and demand} \label{fig:supply} 
\end{figure*}

In Figure \ref{fig:supply}, we show the (real-time) cumulative weekly supply and demand forecasts over time, respectively, in terms of total, A, O, B, and AB CCP units after using MARS for the dataset considering only the available data up to that week. The real-time demand forecasting for each week in Figure \ref{fig:supply} uses the cumulative aggregate demand data over all hospital hubs up to that week. We observe that MARS performs well even with our limited dataset. Table \ref{tab:3} shows Root Mean Square Error (RMSE) and Mean Absolute Percentage Error (MAPE) of our supply and demand forecasts under MARS for total, A, O, B, and AB CCP units; lower values are better. We observe that MARS fits the data well after performing error analysis and accounting for the mentioned challenges discussed in Section \ref{erroranalysis} and Section \ref{challenges}. We note that we are making forecasts from week $4$, which explains the severe supply overestimation for week $4$ in Figure \ref{fig:absupply} where the data is sparse. In such situations, the cumulative value remains the same until a subsequent cumulative forecast is higher.

\begin{table*}[htbp]
\centering
\footnotesize
\begin{tabular}{lllllllllll}
\hline
\multirow{2}{*}{Hub} & \multicolumn{5}{l}{RMSE}  & \multicolumn{5}{l}{MAPE} \\ \cmidrule(lr){2-6} \cmidrule(lr){7-11}   
 & Total & A & O & B & AB & Total (\%) & A (\%) & O (\%) & B (\%) &AB (\%)\\ \hline
Hospital Hub 1 & 1.06 & 0.43 & 0.84 & 0.43 & 0.41 & 16.07 & 17.05 & 19.73 & 20.45 & 22.73 \\
Hospital Hub 2 & 1.74 & 1.21 & 0.91 & 0.84 & 0.58 & 12.87 & 18.79 & 15.50 & 17.82 & 17.86 \\
Hospital Hub 3 & 2.97 & 1.88 & 1.56 & 0.81 & 0.87 & 11.32 & 19.41 & 12.59 & 19.89 & 18.67 \\
Hospital Hub 4 & 2.33 & 1.59 & 0.95 & 0.61 & 0.49 & 21.70 & 25.45 & 53.45 & 16.75 & 16.67 \\
Hospital Hub 5 & 1.76 & 1.10 & 0.80 & 1.02 & 0.46 & 7.31 & 15.26 & 11.52 & 22.35 & 17.36 \\
Hospital Hub 6 & 3.00 & 1.38 & 1.73 & 1.06 & 0.66 & 15.12 & 13.86 & 14.60 & 22.66 & 14.94 \\
Hospital Hub 7 & 2.64 & 2.01 & 1.04 & - & - & 27.91 & 25.58 & 27.12 & - & - \\
All Hospital Hubs & 7.01 & 4.27 & 1.99 & 1.57 & 1.57 & 3.67 & 6.45 & 3.78 & 18.36 & 14.46 \\
CBS Distribution Site & 10.68 & 5.25 & 5.41 & 1.78 & 3.35 & 6.12 & 6.01 & 9.90 & 7.59 & 18.00 \\ \hline
\end{tabular}
\caption{RMSE and MAPE of supply and demand forecasts under MARS}
\label{tab:3}
\end{table*}

All the experiments are run on Spyder Python IDE on a computer with an Intel(R) Core(TM) i7-4510U processor running at $2.60$ GHz using $94$ MB of RAM, running Windows 10 version 21H2.
The weekly supply and demand forecasting took on average approximately $0.04$ and $0.28$ seconds, respectively. Our MIP model has $43$ variables and $81$ constraints, and solving the weekly MIP model took on average approximately $0.02$ seconds. Table \ref{tab:4} reports our CCP allocation model's performance in terms of unmet demand and unmet demand ratio for our primary study, i.e., allocating the resources based on the forecast supply and demand for each blood group. In Table \ref{tab:5}, we compare the results to both when no forecasting is required, i.e., the actual values of supply and demand for each resource are known for each week and the actual allocations in the CONCOR-1 trial. In both tables, two different compatibility matrices (identity and CONCOR-1) are chosen for the MIP allocation model. Finally, in Table \ref{tab:6}, we analyze the sensitivity of our model to different allocation settings by forecasting the aggregate supply and demand and using the distribution of Canadian blood groups to calculate the supply and demand for each resource. A total of $m=300$ runs are considered for calculating unmet demand and unmet demand ratio and the standard error (SE) of their
corresponding $95$\% confidence intervals is reported.

\begin{table*}[htbp]
\centering
\footnotesize   
\begin{tabular}{lllllll}
\hline 
\multirow{2}{*}{} & \multirow{2}{*}{} & \multirow{2}{*}{} & \multicolumn{2}{l}{Identity} & \multicolumn{2}{l}{CONCOR-1} \\ \cmidrule(lr){4-5} \cmidrule(lr){6-7}
{Hub} & {Total Demand} & {Total Forecast Demand} & $\overline{u}_{T}$ & $\overline{z}_{T}$ (\%) & $\overline{u}_{T}$ & $\overline{z}_{T}$ (\%) \\ \hline
Hospital Hub 1 & 12.5 & 12 & 1.00 & 8.00 & 1.00 & 8.00 \\
Hospital Hub 2 & 29 & 30.5 & 0.50 & 1.72 & 1.00 & 3.45 \\
Hospital Hub 3 & 74 & 79 & 4.50 & 6.08 & 10.00 & 13.51 \\
Hospital Hub 4 & 26 & 30.5 & 1.00 & 3.85 & 1.50 & 5.77 \\
Hospital Hub 5 & 30 & 30 & 1.50 & 5.00 & 1.00 & 3.33 \\
Hospital Hub 6 & 105 & 110.5 & 8.50 & 8.10 & 10.00 & 9.52 \\
Hospital Hub 7 & 17.5 & 20.5 & 0 & 0 & 0 & 0 \\ \hline
\end{tabular}
\caption{Resource allocation results under forecast supply and demand for each resource}
\label{tab:4}
\end{table*}

\begin{table*}[htbp]
\centering
\footnotesize   
\begin{tabular}{llllllll}
\hline
\multirow{3}{*}{} & \multirow{3}{*}{} & \multicolumn{4}{l}{MIP Allocations} & {} & {} \\ \cmidrule(lr){3-6}
 {Hub} & {Total Demand} & \multicolumn{2}{l}{Identity} & \multicolumn{2}{l}{CONCOR-1} &  \multicolumn{2}{l}{Allocations in CONCOR-1}  \\ \cmidrule(lr){3-4} \cmidrule(lr){5-6}  \cmidrule(lr){7-8}
 &  & $\overline{u}_{T} $ & $\overline{z}_{T} $ (\%) & $\overline{u}_{T}$ & $\overline{z}_{T}$ (\%)& $\overline{u}_{T}$  & $\overline{z}_{T}$ (\%)  \\ \hline
Hospital Hub 1 & 12.5 & 1.50 & 12.00 & 1.00 & 8.00 & 3.00 & 24.00 \\
Hospital Hub 2 & 29 & 1.00 & 3.45 & 0.50 & 1.72 & 1.50 & 5.17 \\
Hospital Hub 3 & 74 & 7.00 & 9.46 & 4.00 & 5.41 & 20.00 & 27.02 \\
Hospital Hub 4 & 26 & 2.00 & 7.69 & 0.50 & 1.92 & 4.50 & 17.31 \\
Hospital Hub 5 & 30 & 1.50 & 5.00 & 0.50 & 1.67 & 5.00 & 16.67 \\
Hospital Hub 6 & 105 & 10.00 & 9.52 & 7.00 & 6.67 & 27.00 & 25.71 \\
Hospital Hub 7 & 17.5 & 0 & 0 & 0 & 0 & 1.00 & 5.71 \\ \hline
\end{tabular}
\caption{MIP model allocations under actual supply and demand versus actual allocations in CONCOR-1 }
\label{tab:5}
\end{table*}

\begin{table*}[htbp]
\centering
\footnotesize     
\begin{tabular}{lllllll}
\hline 
\multirow{2}{*}{} & \multirow{2}{*}{} & \multirow{2}{*}{} & \multicolumn{2}{l}{Identity} & \multicolumn{2}{l}{CONCOR-1} \\ \cmidrule(lr){4-5} \cmidrule(lr){6-7}
{Hub} & {Total Demand} & {Total Forecast Demand} & $\overline{u}_{T} \pm SE$ & $\overline{z}_{T} \pm SE$ (\%) & $\overline{u}_{T} \pm SE$ & $\overline{z}_{T} \pm SE$ (\%) \\ \hline
Hospital Hub 1 & 12.5 & 10.5 & $1.00 \pm 0.03$ & $8.04 \pm 0.26$ & $0.93 \pm 0.03$ & $7.47 \pm 0.27$ \\
Hospital Hub 2 & 29 & 30 & $1.62 \pm 0.10$ & $5.57 \pm 0.36$ & $1.65 \pm 0.10$ & $5.70 \pm 0.34$ \\
Hospital Hub 3 & 74 & 77 & $3.66 \pm 0.15$ & $4.94 \pm 0.20$ & $5.97 \pm 0.20$ & $8.07 \pm 0.27$ \\
Hospital Hub 4 & 26 & 28.5 & $1.29 \pm 0.08$ & $4.97 \pm 0.31$ & $1.52 \pm 0.08$ & $5.85 \pm 0.30$ \\
Hospital Hub 5 & 30 & 29.5 & $1.26 \pm 0.06$ & $4.20 \pm 0.19$ & $1.21 \pm 0.05$ & $4.03 \pm 0.16$ \\
Hospital Hub 6 & 105 & 110 & $8.57 \pm 0.17$ & $8.16 \pm 0.16$ & $7.63 \pm 0.19$ & $7.27 \pm 0.18$ \\
Hospital Hub 7 & 17.5 & 20.5 & $0.02 \pm 0.01$ & $0.14 \pm 0.07$ & $0.42 \pm 0.03$ & $2.38 \pm 0.15$ \\ \hline
\end{tabular}
\caption{Resource allocation results under aggregate forecast supply and demand (Sensitivity Analysis)}
\label{tab:6}
\end{table*}

The key observations obtained from using our data-driven resource allocation model in the CONCOR-1 case study are as follows:

\begin{legal}[leftmargin=0cm,itemindent=.5cm,labelwidth=\itemindent,labelsep=0cm,listparindent= 0.5cm, align=left]
\item \textbf{The role of hospital hubs’ proportion of CCP demand --- }\emph{A hospital hub's proportion of CCP demand for a resource can lead to its demand not being fully met if the available supply of the resource on a particular week is limited compared to its total demand for the same week. Our approach can help minimize the unmet CCP demand ratios and lead to balanced and fair CCP allocation decisions.} 
\end{legal}

We note that there is a shortage of $10$ O, $5.5$ B, and $7.5$ AB CCP units due to the limitation in supply of these blood groups, as reported in Table \ref{tab:2}. Thus, the CCP demand proportion of a hospital hub for a particular resource and the variance in the total demand between the hospital hubs in each week can affect the unmet demand and unmet demand ratio. For instance, if the available supply on week $t$ can meet all the demand on week $t$, a hospital hub's proportion of CCP demand for a resource is not an issue. On the other hand, if the available supply on week $t$ is limited compared to the total demand for the same week, a hospital hub's high demand proportion for a limited resource CCP can lead to its demand not being fully met. This issue is unavoidable in our setting where we make predictions and allocations on a weekly basis and the optimization model finds a solution based only on the current situation.

We observe in Table \ref{tab:4} that Hospital Hub $1$, Hospital Hub $3$, and Hospital Hub $6$ have larger unmet demand ratios under both compatibility matrices. Hospital Hub $1$'s unmet demand ratio is high because of its unmet CCP demand proportions. However, its unmet demand is not high and is due to the demand that was not met on the last week in our model (and may have been satisfied if we continued for additional weeks). In fact, the supply and demand for the last week can highly affect the final unmet demands of all hospital hubs. Hospital Hub $3$ and Hospital Hub $6$ require a higher proportion of B and AB CCP units compared to the other hospital hubs on weeks when the supply of these units is limited. A hospital hub's unmet demand on a particular week is moved to the next week and will have an effect on its future allocations. We observe that the rest of the hospital hubs have similar unmet demand and unmet demand ratio values. This suggests that our proposed data-driven MIP model leads to a reasonable and fair balance of limited CCP products between the hospital hubs under the MARS forecasting model.

\begin{legal}[leftmargin=0cm,itemindent=.5cm,labelwidth=\itemindent,labelsep=0cm, align=left, resume]
\item \textbf{The role of the compatibility matrix}
\begin{legal} \item \emph{  When the actual supply and demand is unbalanced, not known before allocation, and hence error due to forecasting is present, using the identity compatibility matrix in the MIP level is preferred as it prevents the allocation of limited resources to demand for a more abundant resource.}
\end{legal}

We notice in Table \ref{tab:4} that the total unmet demand for the identity compatibility matrix is lower than for the CONCOR-1 compatibility matrix. While the CONCOR-1 compatibility matrix satisfies as much demand as possible in the current week, its use in this real-time setting can cause issues for future demand, in particular by allocating resources with lower long term supply to demands that can (eventually) be satisfied by more abundant resources. We found that under the CONCOR-1 compatibility matrix, the shortage of O units was compensated by excess A units, and excess B demands were met by AB units. Because the supply of A units was the highest during the trial, and the supply of AB units was the lowest, using AB units to meet demand from other blood groups led to shortages of AB units in future weeks. Furthermore, because the supply and demand forecasts are used instead of the actual values, we might overestimate the demand for a rare resource, such as the AB blood group CCP that is also compatible with another resource, so its allocation increases the final unmet demand of some hospital hubs. However, under the identity compatibility matrix, greedy allocations at the MIP level are prevented. If the forecasting error is not too large, the combination of the identity compatibility matrix at the MIP level and the CONCOR-1 compatibility matrix at the inventory level of hospital hubs is preferred under forecast supply and demand as the identity compatibility matrix at the MIP level prevents greedy allocations that cause issues for future demand, but the forecasting error results in the allocation of some compatible resources that combined with the CONCOR-1 compatibility matrix at the hospital hubs’ inventory level can help meet more demand. Therefore, for our primary study, we recommend this combination of compatibility matrices that prevents future misallocation and absorbs the forecasting errors in an effective manner.

\begin{legal}[resume]
\item \emph{  The CONCOR-1 compatibility matrix results in lower unmet demand when the actual supply and demand for each week is known. The total unmet demand under this compatibility matrix is close to the lowest achievable value, the actual shortage in supply.}
\end{legal}

While using the identity compatibility matrix in Table \ref{tab:4} where supply and demand forecasts are used exhibits better results, the opposite is true when solving the MIP model based on the assumption of knowing the actual supply and demand for each week. This can be observed by comparing the total unmet demand in Table \ref{tab:5} under "MIP Allocations" for both compatibility matrices ($23$ versus $13.5$). When the actual supply and demand are used, over-allocation of limited resources will not happen and the final unmet demand is close to the actual supply shortage. We note again that demand for $5.5$ B and $7.5$ AB CCP units cannot be met due to supply shortage which is almost exactly the total unmet demand result of our MIP model under the CONCOR-1 compatibility matrix ($13.5$). This observation also reinforces the notion in the previous observation that forecasting errors (combined with unbalanced supply and demand) are what drive the recommendation of the identity compatibility matrix under forecast supply and demand. The identity compatibility matrix does not allow cross-transfusion, so $10$ O units cannot be satisfied by the excess A units due to supply shortage. The total unmet demand under the identity compatibility matrix ($23$) further validates the performance of our model in efficiently allocating the available resources as it is equal to the actual shortage in supply, as shown in Table \ref{tab:2}. 

\begin{legal}[resume]
\item \emph{  The identity compatibility matrix results in lower unmet demand under forecast supply and demand compared to when it is used under actual supply and demand, as reasonable forecast error can better help in meeting the demand.}
\end{legal}

Comparing the unmet demand ratio values in Table \ref{tab:4} and Table \ref{tab:5} under the CONCOR-1 compatibility matrix for each hospital hub, we observe that the unmet demand ratio for all hospital hubs is improved (or is the same) in Table \ref{tab:5}, significantly so for Hospital Hub $3$ and Hospital Hub $6$. Furthermore, the unmet demand ratio values in Table \ref{tab:5} are closer to each other under the CONCOR-1 compatibility matrix than those reported in Table \ref{tab:4}, where the results are affected by the forecasting errors. However, we notice that all unmet demand ratio values in Table \ref{tab:4} under the identity compatibility matrix are lower than (or the same as) those in Table \ref{tab:5}. The reason is that using the identity compatibility matrix at the MIP level under actual supply and demand is not a good choice considering that the MIP model never assigns more CCP units than the actual demand. This choice leads to the CONCOR-1 compatibility matrix at the inventory level of the hospital hubs not helping at all in meeting the demand. However, under forecast supply and demand, although the identity compatibility matrix prevents the allocation of a resource to a demand for another resource, the forecasts induce some compatible assignments. Hence, when the identity compatibility matrix at the MIP level is combined with the CONCOR-1 compatibility matrix at the inventory level of the hospital hubs, and the supply and demand forecasts do not have large errors, there is more chance to meet the demand, as what is observed for Hospital Hub $3$ and Hospital Hub $6$.

\begin{legal}[resume]
\item \emph{  The MIP model's results after using the aggregate forecast supply and demand (sensitivity analysis) are close to the primary study when the identity compatibility matrix is used, and improved under the CONCOR-1 compatibility matrix. However, the identity compatibility matrix is a better choice than the CONCOR-1 compatibility matrix when using the aggregate forecast supply and demand and in the presence of forecast errors, consistent with what is observed in the primary study. }
\end{legal}

We notice that although the results under the identity compatibility matrix remain almost unchanged, the total unmet demand and unmet demand ratio values under the CONCOR-1 compatibility matrix in Table \ref{tab:6} are lower than those in Table \ref{tab:4}. This appears to arise due to the fact that including randomness for generating the supply and demand for each blood group based on the aggregate supply and demand may slightly lower the forecasting error. Furthermore, performing sufficient runs helps to smooth the effect of scenarios where resources with limited supply, such as the AB blood group CCP, are allocated in a greedy manner and cause issues for future demand. We note that the results under the identity compatibility matrix are better than those under the CONCOR-1 compatibility matrix in Table \ref{tab:6} as the identity compatibility matrix prevents any cross-allocation of units and lowers the effect of forecast errors, consistent with what we observed in Table \ref{tab:4}. Comparing the unmet demand and unmet demand ratio values under the CONCOR-1 compatibility matrix in Table \ref{tab:5} and Table \ref{tab:6} under "MIP Allocations", it is important to note that while having exact knowledge of supply and demand leads to a fairer allocation, the degradation in performance is not unreasonable if we use the forecast values instead, further supporting the efficacy of our approach. However, as previously discussed, using the identity compatibility matrix in the MIP model under actual supply and demand leads to more unmet demand (as compared to using the CONCOR-1 compatibility matrix) due to not allocating excess units of compatible resources, which prevents the CONCOR-1 compatibility matrix at the inventory level of the hospital hubs to help meet the demand. Similar to our primary study, if the forecast errors are not too large, using the identity compatibility matrix at the MIP level under forecast supply and demand results in making some compatible assignments that can better help meet the demand when combined with the CONCOR-1 compatibility matrix at the hospital hubs’ inventory level.

For this case study and based on all the above observations, we conclude that our primary study when the identity compatibility matrix is used at the MIP level is the most promising choice for our real-time setting. The reason is that as long as the forecast errors for the individual blood groups are not too large, using the combination of the identity compatibility matrix at the MIP level and the CONCOR-1 compatibility matrix at the inventory level helps in both preventing greedy allocations and compensating for forecasting errors. Furthermore, no randomness due to supply and demand probabilities is included in this choice for calculating the supply and demand for each blood group (as what is assumed in our sensitivity analysis), and the assignments of the model are not affected by any ABO blood distributions. 
\end{legal}

\begin{legal}[leftmargin=0cm,itemindent=.5cm,labelwidth=\itemindent,labelsep=0cm,listparindent= 0.5cm, align=left,resume]
\item \textbf{The role of the MIP model --- }\emph{The results obtained from our MIP model both under actual and forecast supply are preferable to what was used in practice.} 
\end{legal}

We compare the unmet demand and unmet demand ratio for both compatibility matrices using the MIP model allocations under actual supply and demand, and the actual allocations in the CONCOR-1 trial. We observe that the results under "MIP Allocations" are more fair than those under "Allocations in CONCOR-1" (the unmet demand ratio values are closer to each other). Furthermore, the total unmet demand under "MIP Allocations" is notably lower ($23$ and $13.5$ versus $62$) which further supports the use of our optimization model, as it accounts for the hospital hubs' unmet demand and inventories.

\section{Conclusion and Future Work} \label{Conclusion}

Decision-makers often face challenges in terms of \begin {enumerate*} \item allocating limited resources, such as vaccines, blood products, and medical equipment, and \item forecasting the supply and demand for these resources during epidemics as there is limited knowledge and historical data about disease demographics. \end {enumerate*}  This work has considered the problem of real-time short-term supply and demand forecasting and fair allocation of limited resources during epidemics by using PLR forecasting models and introducing a data-driven MIP resource allocation model. We have studied the application of our proposed MIP model in a CCP clinical trial case study with the objective of minimizing each hospital hub's unmet ratio of CCP demand. We showed that as long as a hospital hub does not have a high demand proportion for a limited blood group CCP on a particular week (in which case a fair allocation is not possible), our MIP model leads to a balanced and fair final unmet ratio of CCP demand between the hospital hubs. We also showed that allocating a compatible resource to satisfy the demand for a resource helps in situations where the actual supply and demand is known, but might be problematic when we are forecasting the supply and demand and the actual supply of the compatible resource is limited. It would be interesting to investigate the range of the forecasting errors within which a particular compatibility matrix is preferred.

Examining PLR forecasting models on larger datasets and comparing their performance with more advanced machine learning and time-series forecasting models could be investigated in future work. We have addressed several challenges that arise when dealing with sparse data in a real-time setting. We are interested in evaluating other forecasting models that can adapt to these challenges while yielding reasonable forecasts. It would be worthwhile to investigate our MIP model's performance when other supply and demand forecasting models are used.

Finally, multiple objective functions and different notions of fairness can be considered in a single resource allocation problem. Using more than one objective function and focusing on other notions of fairness such as minimizing the aggregate unmet demand over all hospital hubs or minimizing the transportation costs of shipping CCP units with respect to the location of the hospital hubs are examples of problems of interest. It would also be interesting to see our MIP model’s performance when applied to other allocation settings with limited supply. Our MIP model only considers a centralized supplier, thus we leave the model's evaluation with the presence of multiple suppliers to future investigation.

\section*{Acknowledgments}
The authors would like to thank Julie Carruthers, Erin Jamula, and Melanie St John at the McMaster Centre for Transfusion Research for their administrative support.

\section*{Data Availability Statement}
Due to the ethically, legally, and commercially sensitive nature of the research, participants of this study did not agree for their data to be shared publicly, so supporting data is not available.

\section*{Funding}
The authors acknowledge the support and funding provided by a Mitacs Research Training Award (Award IT22358), the McMaster Centre for Transfusion Research, and the Natural Sciences and Engineering Research Council of Canada (NSERC) Discovery Grant Program (RGPIN-2016-04518 and RGPIN-2022-02999).
\section*{Disclosure statement}
No potential competing interest was reported by the authors.

\bibliographystyle{apacite}
\bibliography{references}

\begin{thebibliography}{}

\bibitem [\protect \citeauthoryear {%
Anparasan%
\ \BBA {} Lejeune%
}{%
Anparasan%
\ \BBA {} Lejeune%
}{%
{\protect \APACyear {2019}}%
}]{%
anparasan2019resource}
\APACinsertmetastar {%
anparasan2019resource}%
\begin{APACrefauthors}%
Anparasan, A.%
\BCBT {}\ \BBA {} Lejeune, M.%
\end{APACrefauthors}%
\unskip\
\newblock
\APACrefYearMonthDay{2019}{}{}.
\newblock
{\BBOQ}\APACrefatitle {Resource deployment and donation allocation for epidemic
  outbreaks} {Resource deployment and donation allocation for epidemic
  outbreaks}.{\BBCQ}
\newblock
\APACjournalVolNumPages{Annals of {O}perations {R}esearch}{283}{1}{9--32}.
\PrintBackRefs{\CurrentBib}

\bibitem [\protect \citeauthoryear {%
B{\'e}gin%
\ \protect \BOthers {.}}{%
B{\'e}gin%
\ \protect \BOthers {.}}{%
{\protect \APACyear {2021}}%
}]{%
begin2021convalescent}
\APACinsertmetastar {%
begin2021convalescent}%
\begin{APACrefauthors}%
B{\'e}gin, P.%
, Callum, J.%
, Heddle, N.%
, Cook, R.%
, Zeller, M\BPBI P.%
, Tinmouth, A.%
\BDBL {}Arnold, D\BPBI M.%
\end{APACrefauthors}%
\unskip\
\newblock
\APACrefYearMonthDay{2021}{}{}.
\newblock
{\BBOQ}\APACrefatitle {Convalescent plasma for adults with acute {COVID-19}
  respiratory illness ({CONCOR-1}): {S}tudy protocol for an international,
  multicenter, randomized, open-label trial} {Convalescent plasma for adults
  with acute {COVID-19} respiratory illness ({CONCOR-1}): {S}tudy protocol for
  an international, multicenter, randomized, open-label trial}.{\BBCQ}
\newblock
\APACjournalVolNumPages{{T}rials}{22}{323}{}.
\PrintBackRefs{\CurrentBib}

\bibitem [\protect \citeauthoryear {%
Bekker%
, uit~het Broek%
\BCBL {}\ \BBA {} Koole%
}{%
Bekker%
\ \protect \BOthers {.}}{%
{\protect \APACyear {2023}}%
}]{%
bekker2023modeling}
\APACinsertmetastar {%
bekker2023modeling}%
\begin{APACrefauthors}%
Bekker, R.%
, uit~het Broek, M.%
\BCBL {}\ \BBA {} Koole, G.%
\end{APACrefauthors}%
\unskip\
\newblock
\APACrefYearMonthDay{2023}{}{}.
\newblock
{\BBOQ}\APACrefatitle {Modeling {COVID-19} hospital admissions and occupancy in
  the {N}etherlands} {Modeling {COVID-19} hospital admissions and occupancy in
  the {N}etherlands}.{\BBCQ}
\newblock
\APACjournalVolNumPages{European Journal of Operational
  Research}{304}{1}{207--218}.
\PrintBackRefs{\CurrentBib}

\bibitem [\protect \citeauthoryear {%
Bloch%
\ \protect \BOthers {.}}{%
Bloch%
\ \protect \BOthers {.}}{%
{\protect \APACyear {2021}}%
}]{%
bloch2021guidance}
\APACinsertmetastar {%
bloch2021guidance}%
\begin{APACrefauthors}%
Bloch, E\BPBI M.%
, Goel, R.%
, Wendel, S.%
, Burnouf, T.%
, Al-Riyami, A\BPBI Z.%
, Ang, A\BPBI L.%
\BDBL {}So-Osman, C.%
\end{APACrefauthors}%
\unskip\
\newblock
\APACrefYearMonthDay{2021}{}{}.
\newblock
{\BBOQ}\APACrefatitle {Guidance for the procurement of {COVID-19} convalescent
  plasma: {D}ifferences between high- and low-middle-income countries}
  {Guidance for the procurement of {COVID-19} convalescent plasma:
  {D}ifferences between high- and low-middle-income countries}.{\BBCQ}
\newblock
\APACjournalVolNumPages{Vox {S}anguinis}{116}{1}{18--35}.
\PrintBackRefs{\CurrentBib}

\bibitem [\protect \citeauthoryear {%
Box%
\ \BBA {} Tiao%
}{%
Box%
\ \BBA {} Tiao%
}{%
{\protect \APACyear {1975}}%
}]{%
box1975intervention}
\APACinsertmetastar {%
box1975intervention}%
\begin{APACrefauthors}%
Box, G\BPBI E.%
\BCBT {}\ \BBA {} Tiao, G\BPBI C.%
\end{APACrefauthors}%
\unskip\
\newblock
\APACrefYearMonthDay{1975}{}{}.
\newblock
{\BBOQ}\APACrefatitle {Intervention analysis with applications to economic and
  environmental problems} {Intervention analysis with applications to economic
  and environmental problems}.{\BBCQ}
\newblock
\APACjournalVolNumPages{Journal of the American Statistical
  Association}{70}{349}{70--79}.
\PrintBackRefs{\CurrentBib}

\bibitem [\protect \citeauthoryear {%
Brauer%
}{%
Brauer%
}{%
{\protect \APACyear {2008}}%
}]{%
brauer2008compartmental}
\APACinsertmetastar {%
brauer2008compartmental}%
\begin{APACrefauthors}%
Brauer, F.%
\end{APACrefauthors}%
\unskip\
\newblock
\APACrefYearMonthDay{2008}{}{}.
\newblock
{\BBOQ}\APACrefatitle {Compartmental models in epidemiology} {Compartmental
  models in epidemiology}.{\BBCQ}
\newblock
\BIn{} \APACrefbtitle {Mathematical {E}pidemiology} {Mathematical
  {E}pidemiology}\ (\BPGS\ 19--79).
\newblock
\APACaddressPublisher{}{Springer}.
\PrintBackRefs{\CurrentBib}

\bibitem [\protect \citeauthoryear {%
Butte%
\ \protect \BOthers {.}}{%
Butte%
\ \protect \BOthers {.}}{%
{\protect \APACyear {2010}}%
}]{%
butte2010validation}
\APACinsertmetastar {%
butte2010validation}%
\begin{APACrefauthors}%
Butte, N\BPBI F.%
, Wong, W\BPBI W.%
, Adolph, A\BPBI L.%
, Puyau, M\BPBI R.%
, Vohra, F\BPBI A.%
\BCBL {}\ \BBA {} Zakeri, I\BPBI F.%
\end{APACrefauthors}%
\unskip\
\newblock
\APACrefYearMonthDay{2010}{}{}.
\newblock
{\BBOQ}\APACrefatitle {Validation of cross-sectional time series and
  multivariate adaptive regression splines models for the prediction of energy
  expenditure in children and adolescents using doubly labeled water}
  {Validation of cross-sectional time series and multivariate adaptive
  regression splines models for the prediction of energy expenditure in
  children and adolescents using doubly labeled water}.{\BBCQ}
\newblock
\APACjournalVolNumPages{The {J}ournal of {N}utrition}{140}{8}{1516--1523}.
\PrintBackRefs{\CurrentBib}

\bibitem [\protect \citeauthoryear {%
{Canadian Blood Services}%
}{%
{Canadian Blood Services}%
}{%
{\protect \APACyear {2021}}%
}]{%
whatscbs}
\APACinsertmetastar {%
whatscbs}%
\begin{APACrefauthors}%
{Canadian Blood Services}.%
\end{APACrefauthors}%
\unskip\
\newblock
\APACrefYearMonthDay{2021}{}{}.
\newblock
\APACrefbtitle {{What's my blood type?}} {{What's my blood type?}}
\newblock
\APAChowpublished
  {\url{http://www.blood.ca/en/blood/donating-blood/whats-my-blood-type}}.
\newblock
\APACrefnote{[Online; accessed 20-October-2022]}
\PrintBackRefs{\CurrentBib}

\bibitem [\protect \citeauthoryear {%
Caunhye%
, Nie%
\BCBL {}\ \BBA {} Pokharel%
}{%
Caunhye%
\ \protect \BOthers {.}}{%
{\protect \APACyear {2012}}%
}]{%
caunhye2012optimization}
\APACinsertmetastar {%
caunhye2012optimization}%
\begin{APACrefauthors}%
Caunhye, A\BPBI M.%
, Nie, X.%
\BCBL {}\ \BBA {} Pokharel, S.%
\end{APACrefauthors}%
\unskip\
\newblock
\APACrefYearMonthDay{2012}{}{}.
\newblock
{\BBOQ}\APACrefatitle {Optimization models in emergency logistics: A literature
  review} {Optimization models in emergency logistics: A literature
  review}.{\BBCQ}
\newblock
\APACjournalVolNumPages{Socio-economic Planning Sciences}{46}{1}{4--13}.
\PrintBackRefs{\CurrentBib}

\bibitem [\protect \citeauthoryear {%
Chen%
, Xiong%
, Bao%
\BCBL {}\ \BBA {} Shi%
}{%
Chen%
\ \protect \BOthers {.}}{%
{\protect \APACyear {2020}}%
}]{%
chen2020convalescent}
\APACinsertmetastar {%
chen2020convalescent}%
\begin{APACrefauthors}%
Chen, L.%
, Xiong, J.%
, Bao, L.%
\BCBL {}\ \BBA {} Shi, Y.%
\end{APACrefauthors}%
\unskip\
\newblock
\APACrefYearMonthDay{2020}{}{}.
\newblock
{\BBOQ}\APACrefatitle {Convalescent plasma as a potential therapy for
  {COVID-19}} {Convalescent plasma as a potential therapy for
  {COVID-19}}.{\BBCQ}
\newblock
\APACjournalVolNumPages{The {L}ancet {I}nfectious {D}iseases}{20}{4}{398--400}.
\PrintBackRefs{\CurrentBib}

\bibitem [\protect \citeauthoryear {%
Chou%
, Lee%
, Shao%
\BCBL {}\ \BBA {} Chen%
}{%
Chou%
\ \protect \BOthers {.}}{%
{\protect \APACyear {2004}}%
}]{%
chou2004mining}
\APACinsertmetastar {%
chou2004mining}%
\begin{APACrefauthors}%
Chou, S\BHBI M.%
, Lee, T\BHBI S.%
, Shao, Y\BPBI E.%
\BCBL {}\ \BBA {} Chen, I\BHBI F.%
\end{APACrefauthors}%
\unskip\
\newblock
\APACrefYearMonthDay{2004}{}{}.
\newblock
{\BBOQ}\APACrefatitle {Mining the breast cancer pattern using artificial neural
  networks and multivariate adaptive regression splines} {Mining the breast
  cancer pattern using artificial neural networks and multivariate adaptive
  regression splines}.{\BBCQ}
\newblock
\APACjournalVolNumPages{Expert {S}ystems with {A}pplications}{27}{1}{133--142}.
\PrintBackRefs{\CurrentBib}

\bibitem [\protect \citeauthoryear {%
Chowdhury%
\ \protect \BOthers {.}}{%
Chowdhury%
\ \protect \BOthers {.}}{%
{\protect \APACyear {2021}}%
}]{%
chowdhury2021early}
\APACinsertmetastar {%
chowdhury2021early}%
\begin{APACrefauthors}%
Chowdhury, M\BPBI E.%
, Rahman, T.%
, Khandakar, A.%
, Al-Madeed, S.%
, Zughaier, S\BPBI M.%
, Doi, S\BPBI A\BPBI R.%
\BDBL {}Islam, M\BPBI T.%
\end{APACrefauthors}%
\unskip\
\newblock
\APACrefYearMonthDay{2021}{}{}.
\newblock
{\BBOQ}\APACrefatitle {An early warning tool for predicting mortality risk of
  {COVID-19} patients using machine learning} {An early warning tool for
  predicting mortality risk of {COVID-19} patients using machine
  learning}.{\BBCQ}
\newblock
\APACjournalVolNumPages{Cognitive Computation}{}{}{1--16}.
\PrintBackRefs{\CurrentBib}

\bibitem [\protect \citeauthoryear {%
Dasaklis%
, Pappis%
\BCBL {}\ \BBA {} Rachaniotis%
}{%
Dasaklis%
\ \protect \BOthers {.}}{%
{\protect \APACyear {2012}}%
}]{%
dasaklis2012epidemics}
\APACinsertmetastar {%
dasaklis2012epidemics}%
\begin{APACrefauthors}%
Dasaklis, T\BPBI K.%
, Pappis, C\BPBI P.%
\BCBL {}\ \BBA {} Rachaniotis, N\BPBI P.%
\end{APACrefauthors}%
\unskip\
\newblock
\APACrefYearMonthDay{2012}{}{}.
\newblock
{\BBOQ}\APACrefatitle {Epidemics control and logistics operations: A review}
  {Epidemics control and logistics operations: A review}.{\BBCQ}
\newblock
\APACjournalVolNumPages{International Journal of Production
  Economics}{139}{2}{393--410}.
\PrintBackRefs{\CurrentBib}

\bibitem [\protect \citeauthoryear {%
Du%
, Sai%
\BCBL {}\ \BBA {} Kong%
}{%
Du%
\ \protect \BOthers {.}}{%
{\protect \APACyear {2021}}%
}]{%
du2021data}
\APACinsertmetastar {%
du2021data}%
\begin{APACrefauthors}%
Du, M.%
, Sai, A.%
\BCBL {}\ \BBA {} Kong, N.%
\end{APACrefauthors}%
\unskip\
\newblock
\APACrefYearMonthDay{2021}{}{}.
\newblock
{\BBOQ}\APACrefatitle {A data-driven optimization approach for multi-period
  resource allocation in cholera outbreak control} {A data-driven optimization
  approach for multi-period resource allocation in cholera outbreak
  control}.{\BBCQ}
\newblock
\APACjournalVolNumPages{European {J}ournal of {O}perational
  {R}esearch}{291}{3}{1106--1116}.
\PrintBackRefs{\CurrentBib}

\bibitem [\protect \citeauthoryear {%
Ellaway%
}{%
Ellaway%
}{%
{\protect \APACyear {1978}}%
}]{%
ellaway1978cumulative}
\APACinsertmetastar {%
ellaway1978cumulative}%
\begin{APACrefauthors}%
Ellaway, P.%
\end{APACrefauthors}%
\unskip\
\newblock
\APACrefYearMonthDay{1978}{}{}.
\newblock
{\BBOQ}\APACrefatitle {Cumulative sum technique and its application to the
  analysis of peristimulus time histograms} {Cumulative sum technique and its
  application to the analysis of peristimulus time histograms}.{\BBCQ}
\newblock
\APACjournalVolNumPages{Electroencephalography and {C}linical
  {N}europhysiology}{45}{2}{302--304}.
\PrintBackRefs{\CurrentBib}

\bibitem [\protect \citeauthoryear {%
Fernandez-Sojo%
\ \protect \BOthers {.}}{%
Fernandez-Sojo%
\ \protect \BOthers {.}}{%
{\protect \APACyear {2022}}%
}]{%
fernandez2022hub}
\APACinsertmetastar {%
fernandez2022hub}%
\begin{APACrefauthors}%
Fernandez-Sojo, J.%
, Delgadillo, J.%
, Vives, J.%
, Rodriguez, L.%
, Mendoza, A.%
, Azqueta, C.%
\BDBL {}Querol, S.%
\end{APACrefauthors}%
\unskip\
\newblock
\APACrefYearMonthDay{2022}{}{}.
\newblock
{\BBOQ}\APACrefatitle {A hub-and-spoke model to deliver effective access to
  chimeric antigen receptor {T}-cell therapy in a public health network: {The
  Catalan Blood and Tissue Bank experience}} {A hub-and-spoke model to deliver
  effective access to chimeric antigen receptor {T}-cell therapy in a public
  health network: {The Catalan Blood and Tissue Bank experience}}.{\BBCQ}
\newblock
\APACjournalVolNumPages{Cytotherapy}{}{}{}.
\PrintBackRefs{\CurrentBib}

\bibitem [\protect \citeauthoryear {%
Ferstad%
\ \protect \BOthers {.}}{%
Ferstad%
\ \protect \BOthers {.}}{%
{\protect \APACyear {2020}}%
}]{%
ferstad2020model}
\APACinsertmetastar {%
ferstad2020model}%
\begin{APACrefauthors}%
Ferstad, J\BPBI O.%
, Gu, A\BPBI J.%
, Lee, R\BPBI Y.%
, Thapa, I.%
, Shin, A\BPBI Y.%
, Salomon, J\BPBI A.%
\BDBL {}Scheinker, D.%
\end{APACrefauthors}%
\unskip\
\newblock
\APACrefYearMonthDay{2020}{}{}.
\newblock
{\BBOQ}\APACrefatitle {A model to forecast regional demand for {COVID-19}
  related hospital beds} {A model to forecast regional demand for {COVID-19}
  related hospital beds}.{\BBCQ}
\newblock
\APACjournalVolNumPages{medRxiv}{}{}{}.
\PrintBackRefs{\CurrentBib}

\bibitem [\protect \citeauthoryear {%
Friedman%
}{%
Friedman%
}{%
{\protect \APACyear {1991}}%
}]{%
friedman1991multivariate}
\APACinsertmetastar {%
friedman1991multivariate}%
\begin{APACrefauthors}%
Friedman, J\BPBI H.%
\end{APACrefauthors}%
\unskip\
\newblock
\APACrefYearMonthDay{1991}{}{}.
\newblock
{\BBOQ}\APACrefatitle {Multivariate adaptive regression splines} {Multivariate
  adaptive regression splines}.{\BBCQ}
\newblock
\APACjournalVolNumPages{The {A}nnals of {S}tatistics}{}{}{1--67}.
\PrintBackRefs{\CurrentBib}

\bibitem [\protect \citeauthoryear {%
Gharbharan%
\ \protect \BOthers {.}}{%
Gharbharan%
\ \protect \BOthers {.}}{%
{\protect \APACyear {2020}}%
}]{%
gharbharan2020convalescent}
\APACinsertmetastar {%
gharbharan2020convalescent}%
\begin{APACrefauthors}%
Gharbharan, A.%
, Jordans, C\BPBI C.%
, Geurtsvankessel, C.%
, den Hollander, J\BPBI G.%
, Karim, F.%
, Mollema, F\BPBI P\BPBI N.%
\BDBL {}Rijnders, B\BPBI J.%
\end{APACrefauthors}%
\unskip\
\newblock
\APACrefYearMonthDay{2020}{}{}.
\newblock
{\BBOQ}\APACrefatitle {Convalescent plasma for {COVID-19}. {A} randomized
  clinical trial} {Convalescent plasma for {COVID-19}. {A} randomized clinical
  trial}.{\BBCQ}
\newblock
\APACjournalVolNumPages{MEDRxiv}{}{}{}.
\PrintBackRefs{\CurrentBib}

\bibitem [\protect \citeauthoryear {%
Gong%
\ \protect \BOthers {.}}{%
Gong%
\ \protect \BOthers {.}}{%
{\protect \APACyear {2020}}%
}]{%
gong2020tool}
\APACinsertmetastar {%
gong2020tool}%
\begin{APACrefauthors}%
Gong, J.%
, Ou, J.%
, Qiu, X.%
, Jie, Y.%
, Chen, Y.%
, Yuan, L.%
\BDBL {}Hu, B.%
\end{APACrefauthors}%
\unskip\
\newblock
\APACrefYearMonthDay{2020}{}{}.
\newblock
{\BBOQ}\APACrefatitle {A tool for early prediction of severe coronavirus
  disease 2019 ({COVID-19}): {A} multicenter study using the risk nomogram in
  {W}uhan and {G}uangdong, {C}hina} {A tool for early prediction of severe
  coronavirus disease 2019 ({COVID-19}): {A} multicenter study using the risk
  nomogram in {W}uhan and {G}uangdong, {C}hina}.{\BBCQ}
\newblock
\APACjournalVolNumPages{Clinical {I}nfectious {D}iseases}{71}{15}{833--840}.
\PrintBackRefs{\CurrentBib}

\bibitem [\protect \citeauthoryear {%
Han%
, Preciado%
, Nowzari%
\BCBL {}\ \BBA {} Pappas%
}{%
Han%
\ \protect \BOthers {.}}{%
{\protect \APACyear {2015}}%
}]{%
han2015data}
\APACinsertmetastar {%
han2015data}%
\begin{APACrefauthors}%
Han, S.%
, Preciado, V\BPBI M.%
, Nowzari, C.%
\BCBL {}\ \BBA {} Pappas, G\BPBI J.%
\end{APACrefauthors}%
\unskip\
\newblock
\APACrefYearMonthDay{2015}{}{}.
\newblock
{\BBOQ}\APACrefatitle {Data-driven network resource allocation for controlling
  spreading processes} {Data-driven network resource allocation for controlling
  spreading processes}.{\BBCQ}
\newblock
\APACjournalVolNumPages{{IEEE} {T}ransactions on {N}etwork {S}cience and
  {E}ngineering}{2}{4}{127--138}.
\PrintBackRefs{\CurrentBib}

\bibitem [\protect \citeauthoryear {%
Heitmann%
\ \BBA {} Br{\"u}ggemann%
}{%
Heitmann%
\ \BBA {} Br{\"u}ggemann%
}{%
{\protect \APACyear {2014}}%
}]{%
heitmann2014preference}
\APACinsertmetastar {%
heitmann2014preference}%
\begin{APACrefauthors}%
Heitmann, H.%
\BCBT {}\ \BBA {} Br{\"u}ggemann, W.%
\end{APACrefauthors}%
\unskip\
\newblock
\APACrefYearMonthDay{2014}{}{}.
\newblock
{\BBOQ}\APACrefatitle {Preference-based assignment of university students to
  multiple teaching groups} {Preference-based assignment of university students
  to multiple teaching groups}.{\BBCQ}
\newblock
\APACjournalVolNumPages{OR {S}pectrum}{36}{3}{607--629}.
\PrintBackRefs{\CurrentBib}

\bibitem [\protect \citeauthoryear {%
Hung%
\ \protect \BOthers {.}}{%
Hung%
\ \protect \BOthers {.}}{%
{\protect \APACyear {2011}}%
}]{%
hung2011convalescent}
\APACinsertmetastar {%
hung2011convalescent}%
\begin{APACrefauthors}%
Hung, I\BPBI F.%
, To, K\BPBI K.%
, Lee, C\BHBI K.%
, Lee, K\BHBI L.%
, Chan, K.%
, Yan, W\BHBI W.%
\BDBL {}Yuen, K\BHBI Y.%
\end{APACrefauthors}%
\unskip\
\newblock
\APACrefYearMonthDay{2011}{}{}.
\newblock
{\BBOQ}\APACrefatitle {Convalescent plasma treatment reduced mortality in
  patients with severe pandemic influenza {A} ({H1N1}) $2009$ virus infection}
  {Convalescent plasma treatment reduced mortality in patients with severe
  pandemic influenza {A} ({H1N1}) $2009$ virus infection}.{\BBCQ}
\newblock
\APACjournalVolNumPages{Clinical {I}nfectious {D}iseases}{52}{4}{447--456}.
\PrintBackRefs{\CurrentBib}

\bibitem [\protect \citeauthoryear {%
Jarrett%
\ \BBA {} Khumuwala%
}{%
Jarrett%
\ \BBA {} Khumuwala%
}{%
{\protect \APACyear {1987}}%
}]{%
jarrett1987study}
\APACinsertmetastar {%
jarrett1987study}%
\begin{APACrefauthors}%
Jarrett, J\BPBI E.%
\BCBT {}\ \BBA {} Khumuwala, S\BPBI B.%
\end{APACrefauthors}%
\unskip\
\newblock
\APACrefYearMonthDay{1987}{}{}.
\newblock
{\BBOQ}\APACrefatitle {A study of forecast error and covariant time series to
  improve forecasting for financial decision making} {A study of forecast error
  and covariant time series to improve forecasting for financial decision
  making}.{\BBCQ}
\newblock
\APACjournalVolNumPages{Managerial {F}inance}{13}{2}{20--24}.
\PrintBackRefs{\CurrentBib}

\bibitem [\protect \citeauthoryear {%
Karsu%
\ \BBA {} Erkan%
}{%
Karsu%
\ \BBA {} Erkan%
}{%
{\protect \APACyear {2020}}%
}]{%
karsu2020balance}
\APACinsertmetastar {%
karsu2020balance}%
\begin{APACrefauthors}%
Karsu, {\"O}.%
\BCBT {}\ \BBA {} Erkan, H.%
\end{APACrefauthors}%
\unskip\
\newblock
\APACrefYearMonthDay{2020}{}{}.
\newblock
{\BBOQ}\APACrefatitle {Balance in resource allocation problems: {A} changing
  reference approach} {Balance in resource allocation problems: {A} changing
  reference approach}.{\BBCQ}
\newblock
\APACjournalVolNumPages{{OR} {S}pectrum}{42}{1}{297--326}.
\PrintBackRefs{\CurrentBib}

\bibitem [\protect \citeauthoryear {%
Karsu%
\ \BBA {} Morton%
}{%
Karsu%
\ \BBA {} Morton%
}{%
{\protect \APACyear {2014}}%
}]{%
karsu2014incorporating}
\APACinsertmetastar {%
karsu2014incorporating}%
\begin{APACrefauthors}%
Karsu, {\"O}.%
\BCBT {}\ \BBA {} Morton, A.%
\end{APACrefauthors}%
\unskip\
\newblock
\APACrefYearMonthDay{2014}{}{}.
\newblock
{\BBOQ}\APACrefatitle {Incorporating balance concerns in resource allocation
  decisions: {A} bi-criteria modelling approach} {Incorporating balance
  concerns in resource allocation decisions: {A} bi-criteria modelling
  approach}.{\BBCQ}
\newblock
\APACjournalVolNumPages{Omega}{44}{}{70--82}.
\PrintBackRefs{\CurrentBib}

\bibitem [\protect \citeauthoryear {%
Karsu%
\ \BBA {} Morton%
}{%
Karsu%
\ \BBA {} Morton%
}{%
{\protect \APACyear {2015}}%
}]{%
karsu2015inequity}
\APACinsertmetastar {%
karsu2015inequity}%
\begin{APACrefauthors}%
Karsu, {\"O}.%
\BCBT {}\ \BBA {} Morton, A.%
\end{APACrefauthors}%
\unskip\
\newblock
\APACrefYearMonthDay{2015}{}{}.
\newblock
{\BBOQ}\APACrefatitle {Inequity averse optimization in operational research}
  {Inequity averse optimization in operational research}.{\BBCQ}
\newblock
\APACjournalVolNumPages{European {J}ournal of {O}perational
  {R}esearch}{245}{2}{343--359}.
\PrintBackRefs{\CurrentBib}

\bibitem [\protect \citeauthoryear {%
Katris%
}{%
Katris%
}{%
{\protect \APACyear {2021}}%
}]{%
katris2021time}
\APACinsertmetastar {%
katris2021time}%
\begin{APACrefauthors}%
Katris, C.%
\end{APACrefauthors}%
\unskip\
\newblock
\APACrefYearMonthDay{2021}{}{}.
\newblock
{\BBOQ}\APACrefatitle {A time series-based statistical approach for outbreak
  spread forecasting: {A}pplication of {COVID-19} in {G}reece} {A time
  series-based statistical approach for outbreak spread forecasting:
  {A}pplication of {COVID-19} in {G}reece}.{\BBCQ}
\newblock
\APACjournalVolNumPages{Expert {S}ystems with {A}pplications}{166}{}{114077}.
\PrintBackRefs{\CurrentBib}

\bibitem [\protect \citeauthoryear {%
Kraft%
\ \protect \BOthers {.}}{%
Kraft%
\ \protect \BOthers {.}}{%
{\protect \APACyear {2015}}%
}]{%
kraft2015use}
\APACinsertmetastar {%
kraft2015use}%
\begin{APACrefauthors}%
Kraft, C\BPBI S.%
, Hewlett, A\BPBI L.%
, Koepsell, S.%
, Winkler, A\BPBI M.%
, Kratochvil, C\BPBI J.%
, Larson, L.%
\BDBL {}Ribner, B\BPBI S.%
\end{APACrefauthors}%
\unskip\
\newblock
\APACrefYearMonthDay{2015}{}{}.
\newblock
{\BBOQ}\APACrefatitle {The use of {TKM-100802} and convalescent plasma in $2$
  patients with {E}bola virus disease in the {U}nited {S}tates} {The use of
  {TKM-100802} and convalescent plasma in $2$ patients with {E}bola virus
  disease in the {U}nited {S}tates}.{\BBCQ}
\newblock
\APACjournalVolNumPages{Clinical {I}nfectious {D}iseases}{61}{4}{496--502}.
\PrintBackRefs{\CurrentBib}

\bibitem [\protect \citeauthoryear {%
Kumar%
\ \BBA {} Kleinberg%
}{%
Kumar%
\ \BBA {} Kleinberg%
}{%
{\protect \APACyear {2000}}%
}]{%
kumar2000fairness}
\APACinsertmetastar {%
kumar2000fairness}%
\begin{APACrefauthors}%
Kumar, A.%
\BCBT {}\ \BBA {} Kleinberg, J.%
\end{APACrefauthors}%
\unskip\
\newblock
\APACrefYearMonthDay{2000}{}{}.
\newblock
{\BBOQ}\APACrefatitle {Fairness measures for resource allocation} {Fairness
  measures for resource allocation}.{\BBCQ}
\newblock
\BIn{} \APACrefbtitle {Proceedings 41st {A}nnual {S}ymposium on {F}oundations
  of {C}omputer {S}cience} {Proceedings 41st {A}nnual {S}ymposium on
  {F}oundations of {C}omputer {S}cience}\ (\BPGS\ 75--85).
\PrintBackRefs{\CurrentBib}

\bibitem [\protect \citeauthoryear {%
Lasheras%
\ \protect \BOthers {.}}{%
Lasheras%
\ \protect \BOthers {.}}{%
{\protect \APACyear {2020}}%
}]{%
lasheras2020methodology}
\APACinsertmetastar {%
lasheras2020methodology}%
\begin{APACrefauthors}%
Lasheras, J\BPBI E\BPBI S.%
, Donquiles, C\BPBI G.%
, Nieto, P\BPBI J\BPBI G.%
, Moleon, J\BPBI J\BPBI J.%
, Salas, D.%
, G{\'o}mez, S\BPBI L\BPBI S.%
\BDBL {}de Cos~Juez, F\BPBI J.%
\end{APACrefauthors}%
\unskip\
\newblock
\APACrefYearMonthDay{2020}{}{}.
\newblock
{\BBOQ}\APACrefatitle {A methodology for detecting relevant single nucleotide
  polymorphism in prostate cancer with multivariate adaptive regression splines
  and backpropagation artificial neural networks} {A methodology for detecting
  relevant single nucleotide polymorphism in prostate cancer with multivariate
  adaptive regression splines and backpropagation artificial neural
  networks}.{\BBCQ}
\newblock
\APACjournalVolNumPages{Neural {C}omputing and
  {A}pplications}{32}{5}{1231--1238}.
\PrintBackRefs{\CurrentBib}

\bibitem [\protect \citeauthoryear {%
L.~Li%
\ \protect \BOthers {.}}{%
L.~Li%
\ \protect \BOthers {.}}{%
{\protect \APACyear {2020}}%
}]{%
li2020effect}
\APACinsertmetastar {%
li2020effect}%
\begin{APACrefauthors}%
Li, L.%
, Zhang, W.%
, Hu, Y.%
, Tong, X.%
, Zheng, S.%
, Yang, J.%
\BDBL {}Liu, Z.%
\end{APACrefauthors}%
\unskip\
\newblock
\APACrefYearMonthDay{2020}{}{}.
\newblock
{\BBOQ}\APACrefatitle {Effect of convalescent plasma therapy on time to
  clinical improvement in patients with severe and life-threatening {COVID-19}:
  a randomized clinical trial} {Effect of convalescent plasma therapy on time
  to clinical improvement in patients with severe and life-threatening
  {COVID-19}: a randomized clinical trial}.{\BBCQ}
\newblock
\APACjournalVolNumPages{Jama}{324}{5}{460--470}.
\PrintBackRefs{\CurrentBib}

\bibitem [\protect \citeauthoryear {%
N.~Li%
, Chiang%
, Down%
\BCBL {}\ \BBA {} Heddle%
}{%
N.~Li%
\ \protect \BOthers {.}}{%
{\protect \APACyear {2021}}%
}]{%
li2021decision}
\APACinsertmetastar {%
li2021decision}%
\begin{APACrefauthors}%
Li, N.%
, Chiang, F.%
, Down, D\BPBI G.%
\BCBL {}\ \BBA {} Heddle, N\BPBI M.%
\end{APACrefauthors}%
\unskip\
\newblock
\APACrefYearMonthDay{2021}{}{}.
\newblock
{\BBOQ}\APACrefatitle {A decision integration strategy for short-term demand
  forecasting and ordering for red blood cell components} {A decision
  integration strategy for short-term demand forecasting and ordering for red
  blood cell components}.{\BBCQ}
\newblock
\APACjournalVolNumPages{Operations {R}esearch for {H}ealth {C}are}{}{}{100290}.
\PrintBackRefs{\CurrentBib}

\bibitem [\protect \citeauthoryear {%
N.~Li%
\ \protect \BOthers {.}}{%
N.~Li%
\ \protect \BOthers {.}}{%
{\protect \APACyear {2022}}%
}]{%
li2022data}
\APACinsertmetastar {%
li2022data}%
\begin{APACrefauthors}%
Li, N.%
, Zeller, M\BPBI P.%
, Shih, A\BPBI W.%
, Heddle, N\BPBI M.%
, St.~John, M.%
, B{\'e}gin, P.%
\BDBL {}Tinmouth, A.%
\end{APACrefauthors}%
\unskip\
\newblock
\APACrefYearMonthDay{2022}{}{}.
\newblock
{\BBOQ}\APACrefatitle {A data-informed system to manage scarce blood product
  allocation in a randomized controlled trial of convalescent plasma} {A
  data-informed system to manage scarce blood product allocation in a
  randomized controlled trial of convalescent plasma}.{\BBCQ}
\newblock
\APACjournalVolNumPages{Transfusion}{}{}{}.
\newblock
\begin{APACrefURL}
  \url{https://onlinelibrary.wiley.com/doi/abs/10.1111/trf.17151}
  \end{APACrefURL}
\newblock
\begin{APACrefDOI} \doi{https://doi.org/10.1111/trf.17151} \end{APACrefDOI}
\PrintBackRefs{\CurrentBib}

\bibitem [\protect \citeauthoryear {%
Liu%
\ \BBA {} Liang%
}{%
Liu%
\ \BBA {} Liang%
}{%
{\protect \APACyear {2013}}%
}]{%
liu2013dynamic}
\APACinsertmetastar {%
liu2013dynamic}%
\begin{APACrefauthors}%
Liu, M.%
\BCBT {}\ \BBA {} Liang, J.%
\end{APACrefauthors}%
\unskip\
\newblock
\APACrefYearMonthDay{2013}{}{}.
\newblock
{\BBOQ}\APACrefatitle {Dynamic optimization model for allocating medical
  resources in epidemic controlling} {Dynamic optimization model for allocating
  medical resources in epidemic controlling}.{\BBCQ}
\newblock
\APACjournalVolNumPages{Journal of {I}ndustrial {E}ngineering and {M}anagement
  ({JIEM})}{6}{1}{73--88}.
\PrintBackRefs{\CurrentBib}

\bibitem [\protect \citeauthoryear {%
L{\'o}pez-Lozano%
\ \protect \BOthers {.}}{%
L{\'o}pez-Lozano%
\ \protect \BOthers {.}}{%
{\protect \APACyear {2019}}%
}]{%
lopez2019nonlinear}
\APACinsertmetastar {%
lopez2019nonlinear}%
\begin{APACrefauthors}%
L{\'o}pez-Lozano, J\BHBI M.%
, Lawes, T.%
, Nebot, C.%
, Beyaert, A.%
, Bertrand, X.%
, Hocquet, D.%
\BDBL {}study group, T.%
\end{APACrefauthors}%
\unskip\
\newblock
\APACrefYearMonthDay{2019}{}{}.
\newblock
{\BBOQ}\APACrefatitle {A nonlinear time-series analysis approach to identify
  thresholds in associations between population antibiotic use and rates of
  resistance} {A nonlinear time-series analysis approach to identify thresholds
  in associations between population antibiotic use and rates of
  resistance}.{\BBCQ}
\newblock
\APACjournalVolNumPages{Nature {M}icrobiology}{4}{7}{1160--1172}.
\PrintBackRefs{\CurrentBib}

\bibitem [\protect \citeauthoryear {%
Lu%
\ \BBA {} Chen%
}{%
Lu%
\ \BBA {} Chen%
}{%
{\protect \APACyear {2020}}%
}]{%
lu2020improved}
\APACinsertmetastar {%
lu2020improved}%
\begin{APACrefauthors}%
Lu, M.%
\BCBT {}\ \BBA {} Chen, Y.%
\end{APACrefauthors}%
\unskip\
\newblock
\APACrefYearMonthDay{2020}{}{}.
\newblock
{\BBOQ}\APACrefatitle {Improved estimation and forecasting through
  residual-based model error quantification} {Improved estimation and
  forecasting through residual-based model error quantification}.{\BBCQ}
\newblock
\APACjournalVolNumPages{{SPE} {J}ournal}{25}{02}{951--968}.
\PrintBackRefs{\CurrentBib}

\bibitem [\protect \citeauthoryear {%
Mestre%
, Oliveira%
\BCBL {}\ \BBA {} Barbosa-P{\'o}voa%
}{%
Mestre%
\ \protect \BOthers {.}}{%
{\protect \APACyear {2012}}%
}]{%
mestre2012organizing}
\APACinsertmetastar {%
mestre2012organizing}%
\begin{APACrefauthors}%
Mestre, A\BPBI M.%
, Oliveira, M\BPBI D.%
\BCBL {}\ \BBA {} Barbosa-P{\'o}voa, A.%
\end{APACrefauthors}%
\unskip\
\newblock
\APACrefYearMonthDay{2012}{}{}.
\newblock
{\BBOQ}\APACrefatitle {Organizing hospitals into networks: {A} hierarchical and
  multiservice model to define location, supply and referrals in planned
  hospital systems} {Organizing hospitals into networks: {A} hierarchical and
  multiservice model to define location, supply and referrals in planned
  hospital systems}.{\BBCQ}
\newblock
\APACjournalVolNumPages{{OR} {S}pectrum}{34}{2}{319--348}.
\PrintBackRefs{\CurrentBib}

\bibitem [\protect \citeauthoryear {%
Nikolopoulos%
, Punia%
, Sch{\"a}fers%
, Tsinopoulos%
\BCBL {}\ \BBA {} Vasilakis%
}{%
Nikolopoulos%
\ \protect \BOthers {.}}{%
{\protect \APACyear {2021}}%
}]{%
nikolopoulos2021forecasting}
\APACinsertmetastar {%
nikolopoulos2021forecasting}%
\begin{APACrefauthors}%
Nikolopoulos, K.%
, Punia, S.%
, Sch{\"a}fers, A.%
, Tsinopoulos, C.%
\BCBL {}\ \BBA {} Vasilakis, C.%
\end{APACrefauthors}%
\unskip\
\newblock
\APACrefYearMonthDay{2021}{}{}.
\newblock
{\BBOQ}\APACrefatitle {Forecasting and planning during a pandemic: {COVID-19}
  growth rates, supply chain disruptions, and governmental decisions}
  {Forecasting and planning during a pandemic: {COVID-19} growth rates, supply
  chain disruptions, and governmental decisions}.{\BBCQ}
\newblock
\APACjournalVolNumPages{European {J}ournal of {O}perational
  {R}esearch}{290}{1}{99--115}.
\PrintBackRefs{\CurrentBib}

\bibitem [\protect \citeauthoryear {%
Piantadosi%
}{%
Piantadosi%
}{%
{\protect \APACyear {2017}}%
}]{%
piantadosi2017clinical}
\APACinsertmetastar {%
piantadosi2017clinical}%
\begin{APACrefauthors}%
Piantadosi, S.%
\end{APACrefauthors}%
\unskip\
\newblock
\APACrefYear{2017}.
\newblock
\APACrefbtitle {Clinical trials: A Methodologic Perspective} {Clinical trials:
  A methodologic perspective}.
\newblock
\APACaddressPublisher{}{John Wiley \& Sons}.
\PrintBackRefs{\CurrentBib}

\bibitem [\protect \citeauthoryear {%
Preciado%
, Zargham%
, Enyioha%
, Jadbabaie%
\BCBL {}\ \BBA {} Pappas%
}{%
Preciado%
\ \protect \BOthers {.}}{%
{\protect \APACyear {2013}}%
}]{%
preciado2013optimal}
\APACinsertmetastar {%
preciado2013optimal}%
\begin{APACrefauthors}%
Preciado, V\BPBI M.%
, Zargham, M.%
, Enyioha, C.%
, Jadbabaie, A.%
\BCBL {}\ \BBA {} Pappas, G.%
\end{APACrefauthors}%
\unskip\
\newblock
\APACrefYearMonthDay{2013}{}{}.
\newblock
{\BBOQ}\APACrefatitle {Optimal vaccine allocation to control epidemic outbreaks
  in arbitrary networks} {Optimal vaccine allocation to control epidemic
  outbreaks in arbitrary networks}.{\BBCQ}
\newblock
\BIn{} \APACrefbtitle {52nd IEEE {C}onference on {D}ecision and {C}ontrol}
  {52nd ieee {C}onference on {D}ecision and {C}ontrol}\ (\BPGS\ 7486--7491).
\PrintBackRefs{\CurrentBib}

\bibitem [\protect \citeauthoryear {%
Preciado%
, Zargham%
, Enyioha%
, Jadbabaie%
\BCBL {}\ \BBA {} Pappas%
}{%
Preciado%
\ \protect \BOthers {.}}{%
{\protect \APACyear {2014}}%
}]{%
preciado2014optimal}
\APACinsertmetastar {%
preciado2014optimal}%
\begin{APACrefauthors}%
Preciado, V\BPBI M.%
, Zargham, M.%
, Enyioha, C.%
, Jadbabaie, A.%
\BCBL {}\ \BBA {} Pappas, G\BPBI J.%
\end{APACrefauthors}%
\unskip\
\newblock
\APACrefYearMonthDay{2014}{}{}.
\newblock
{\BBOQ}\APACrefatitle {Optimal resource allocation for network protection
  against spreading processes} {Optimal resource allocation for network
  protection against spreading processes}.{\BBCQ}
\newblock
\APACjournalVolNumPages{{IEEE} {T}ransactions on {C}ontrol of {N}etwork
  {S}ystems}{1}{1}{99--108}.
\PrintBackRefs{\CurrentBib}

\bibitem [\protect \citeauthoryear {%
Psaraftis%
, Tharakan%
\BCBL {}\ \BBA {} Ceder%
}{%
Psaraftis%
\ \protect \BOthers {.}}{%
{\protect \APACyear {1986}}%
}]{%
psaraftis1986optimal}
\APACinsertmetastar {%
psaraftis1986optimal}%
\begin{APACrefauthors}%
Psaraftis, H\BPBI N.%
, Tharakan, G\BPBI G.%
\BCBL {}\ \BBA {} Ceder, A.%
\end{APACrefauthors}%
\unskip\
\newblock
\APACrefYearMonthDay{1986}{}{}.
\newblock
{\BBOQ}\APACrefatitle {Optimal response to oil spills: The strategic decision
  case} {Optimal response to oil spills: The strategic decision case}.{\BBCQ}
\newblock
\APACjournalVolNumPages{Operations Research}{34}{2}{203--217}.
\PrintBackRefs{\CurrentBib}

\bibitem [\protect \citeauthoryear {%
N.~Rachaniotis%
, Dasaklis%
\BCBL {}\ \BBA {} Pappis%
}{%
N.~Rachaniotis%
\ \protect \BOthers {.}}{%
{\protect \APACyear {2017}}%
}]{%
rachaniotis2017controlling}
\APACinsertmetastar {%
rachaniotis2017controlling}%
\begin{APACrefauthors}%
Rachaniotis, N.%
, Dasaklis, T\BPBI K.%
\BCBL {}\ \BBA {} Pappis, C.%
\end{APACrefauthors}%
\unskip\
\newblock
\APACrefYearMonthDay{2017}{}{}.
\newblock
{\BBOQ}\APACrefatitle {Controlling infectious disease outbreaks: {A}
  deterministic allocation-scheduling model with multiple discrete resources}
  {Controlling infectious disease outbreaks: {A} deterministic
  allocation-scheduling model with multiple discrete resources}.{\BBCQ}
\newblock
\APACjournalVolNumPages{Journal of {S}ystems {S}cience and {S}ystems
  {E}ngineering}{26}{2}{219--239}.
\PrintBackRefs{\CurrentBib}

\bibitem [\protect \citeauthoryear {%
N\BPBI P.~Rachaniotis%
, Dasaklis%
\BCBL {}\ \BBA {} Pappis%
}{%
N\BPBI P.~Rachaniotis%
\ \protect \BOthers {.}}{%
{\protect \APACyear {2012}}%
}]{%
rachaniotis2012deterministic}
\APACinsertmetastar {%
rachaniotis2012deterministic}%
\begin{APACrefauthors}%
Rachaniotis, N\BPBI P.%
, Dasaklis, T\BPBI K.%
\BCBL {}\ \BBA {} Pappis, C\BPBI P.%
\end{APACrefauthors}%
\unskip\
\newblock
\APACrefYearMonthDay{2012}{}{}.
\newblock
{\BBOQ}\APACrefatitle {A deterministic resource scheduling model in epidemic
  control: {A} case study} {A deterministic resource scheduling model in
  epidemic control: {A} case study}.{\BBCQ}
\newblock
\APACjournalVolNumPages{European {J}ournal of {O}perational
  {R}esearch}{216}{1}{225--231}.
\PrintBackRefs{\CurrentBib}

\bibitem [\protect \citeauthoryear {%
Roberts%
\ \protect \BOthers {.}}{%
Roberts%
\ \protect \BOthers {.}}{%
{\protect \APACyear {2018}}%
}]{%
roberts2018evaluation}
\APACinsertmetastar {%
roberts2018evaluation}%
\begin{APACrefauthors}%
Roberts, H\BPBI W.%
, Wagh, V\BPBI K.%
, Mullens, I\BPBI J.%
, Borsci, S.%
, Ni, M\BPBI Z.%
\BCBL {}\ \BBA {} O’Brart, D\BPBI P.%
\end{APACrefauthors}%
\unskip\
\newblock
\APACrefYearMonthDay{2018}{}{}.
\newblock
{\BBOQ}\APACrefatitle {Evaluation of a hub-and-spoke model for the delivery of
  femtosecond laser-assisted cataract surgery within the context of a large
  randomised controlled trial} {Evaluation of a hub-and-spoke model for the
  delivery of femtosecond laser-assisted cataract surgery within the context of
  a large randomised controlled trial}.{\BBCQ}
\newblock
\APACjournalVolNumPages{British Journal of Ophthalmology}{102}{11}{1556--1563}.
\PrintBackRefs{\CurrentBib}

\bibitem [\protect \citeauthoryear {%
Rosen%
\ \BBA {} Pardalos%
}{%
Rosen%
\ \BBA {} Pardalos%
}{%
{\protect \APACyear {1986}}%
}]{%
rosen1986global}
\APACinsertmetastar {%
rosen1986global}%
\begin{APACrefauthors}%
Rosen, J\BPBI B.%
\BCBT {}\ \BBA {} Pardalos, P\BPBI M.%
\end{APACrefauthors}%
\unskip\
\newblock
\APACrefYearMonthDay{1986}{}{}.
\newblock
{\BBOQ}\APACrefatitle {Global minimization of large-scale constrained concave
  quadratic problems by separable programming} {Global minimization of
  large-scale constrained concave quadratic problems by separable
  programming}.{\BBCQ}
\newblock
\APACjournalVolNumPages{Mathematical {P}rogramming}{34}{2}{163--174}.
\PrintBackRefs{\CurrentBib}

\bibitem [\protect \citeauthoryear {%
Rudy%
}{%
Rudy%
}{%
{\protect \APACyear {2016}}%
}]{%
rudy2016py}
\APACinsertmetastar {%
rudy2016py}%
\begin{APACrefauthors}%
Rudy, J.%
\end{APACrefauthors}%
\unskip\
\newblock
\APACrefYearMonthDay{2016}{}{}.
\newblock
\APACrefbtitle {py-earth: {A} {P}ython implementation of {J}erome
  {F}riedman’s multivariate adaptive regression splines.} {py-earth: {A}
  {P}ython implementation of {J}erome {F}riedman’s multivariate adaptive
  regression splines.}
\newblock
\APAChowpublished {\url{http://www.github.com/scikit-learn-contrib/py-earth}}.
\newblock
\APACrefnote{[Online; accessed 20-October-2022]}
\PrintBackRefs{\CurrentBib}

\bibitem [\protect \citeauthoryear {%
Senthilkumar%
\ \BBA {} Paulraj%
}{%
Senthilkumar%
\ \BBA {} Paulraj%
}{%
{\protect \APACyear {2013}}%
}]{%
senthilkumar2013diabetes}
\APACinsertmetastar {%
senthilkumar2013diabetes}%
\begin{APACrefauthors}%
Senthilkumar, D.%
\BCBT {}\ \BBA {} Paulraj, S.%
\end{APACrefauthors}%
\unskip\
\newblock
\APACrefYearMonthDay{2013}{}{}.
\newblock
{\BBOQ}\APACrefatitle {Diabetes disease diagnosis using multivariate adaptive
  regression splines} {Diabetes disease diagnosis using multivariate adaptive
  regression splines}.{\BBCQ}
\newblock
\APACjournalVolNumPages{{AGE}}{768}{}{52}.
\PrintBackRefs{\CurrentBib}

\bibitem [\protect \citeauthoryear {%
Serrano%
, S{\'a}nchez%
, Lasheras%
, Iglesias-Rodr{\'\i}guez%
\BCBL {}\ \BBA {} Valverde%
}{%
Serrano%
\ \protect \BOthers {.}}{%
{\protect \APACyear {2020}}%
}]{%
serrano2020identification}
\APACinsertmetastar {%
serrano2020identification}%
\begin{APACrefauthors}%
Serrano, N\BPBI B.%
, S{\'a}nchez, A\BPBI S.%
, Lasheras, F\BPBI S.%
, Iglesias-Rodr{\'\i}guez, F\BPBI J.%
\BCBL {}\ \BBA {} Valverde, G\BPBI F.%
\end{APACrefauthors}%
\unskip\
\newblock
\APACrefYearMonthDay{2020}{}{}.
\newblock
{\BBOQ}\APACrefatitle {Identification of gender differences in the factors
  influencing shoulders, neck and upper limb {MSD} by means of multivariate
  adaptive regression splines ({MARS})} {Identification of gender differences
  in the factors influencing shoulders, neck and upper limb {MSD} by means of
  multivariate adaptive regression splines ({MARS})}.{\BBCQ}
\newblock
\APACjournalVolNumPages{Applied {E}rgonomics}{82}{}{102981}.
\PrintBackRefs{\CurrentBib}

\bibitem [\protect \citeauthoryear {%
Silal%
, Little%
, Barnes%
\BCBL {}\ \BBA {} White%
}{%
Silal%
\ \protect \BOthers {.}}{%
{\protect \APACyear {2016}}%
}]{%
silal2016sensitivity}
\APACinsertmetastar {%
silal2016sensitivity}%
\begin{APACrefauthors}%
Silal, S\BPBI P.%
, Little, F.%
, Barnes, K\BPBI I.%
\BCBL {}\ \BBA {} White, L\BPBI J.%
\end{APACrefauthors}%
\unskip\
\newblock
\APACrefYearMonthDay{2016}{}{}.
\newblock
{\BBOQ}\APACrefatitle {Sensitivity to model structure: {A} comparison of
  compartmental models in epidemiology} {Sensitivity to model structure: {A}
  comparison of compartmental models in epidemiology}.{\BBCQ}
\newblock
\APACjournalVolNumPages{Health {S}ystems}{5}{3}{178--191}.
\PrintBackRefs{\CurrentBib}

\bibitem [\protect \citeauthoryear {%
Simonovich%
\ \protect \BOthers {.}}{%
Simonovich%
\ \protect \BOthers {.}}{%
{\protect \APACyear {2021}}%
}]{%
simonovich2021randomized}
\APACinsertmetastar {%
simonovich2021randomized}%
\begin{APACrefauthors}%
Simonovich, V\BPBI A.%
, Burgos~Pratx, L\BPBI D.%
, Scibona, P.%
, Beruto, M\BPBI V.%
, Vallone, M\BPBI G.%
, V{\'a}zquez, C.%
\BDBL {}Belloso, W\BPBI H.%
\end{APACrefauthors}%
\unskip\
\newblock
\APACrefYearMonthDay{2021}{}{}.
\newblock
{\BBOQ}\APACrefatitle {A randomized trial of convalescent plasma in {Covid-19}
  severe pneumonia} {A randomized trial of convalescent plasma in {Covid-19}
  severe pneumonia}.{\BBCQ}
\newblock
\APACjournalVolNumPages{New England Journal of Medicine}{384}{7}{619--629}.
\PrintBackRefs{\CurrentBib}

\bibitem [\protect \citeauthoryear {%
Smith%
, Harper%
\BCBL {}\ \BBA {} Potts%
}{%
Smith%
\ \protect \BOthers {.}}{%
{\protect \APACyear {2013}}%
}]{%
smith2013bicriteria}
\APACinsertmetastar {%
smith2013bicriteria}%
\begin{APACrefauthors}%
Smith, H\BPBI K.%
, Harper, P\BPBI R.%
\BCBL {}\ \BBA {} Potts, C\BPBI N.%
\end{APACrefauthors}%
\unskip\
\newblock
\APACrefYearMonthDay{2013}{}{}.
\newblock
{\BBOQ}\APACrefatitle {Bicriteria efficiency/equity hierarchical location
  models for public service application} {Bicriteria efficiency/equity
  hierarchical location models for public service application}.{\BBCQ}
\newblock
\APACjournalVolNumPages{{J}ournal of the {O}perational {R}esearch
  {S}ociety}{64}{4}{500--512}.
\PrintBackRefs{\CurrentBib}

\bibitem [\protect \citeauthoryear {%
Srinivasa%
\ \BBA {} Wilhelm%
}{%
Srinivasa%
\ \BBA {} Wilhelm%
}{%
{\protect \APACyear {1997}}%
}]{%
srinivasa1997procedure}
\APACinsertmetastar {%
srinivasa1997procedure}%
\begin{APACrefauthors}%
Srinivasa, A\BPBI V.%
\BCBT {}\ \BBA {} Wilhelm, W\BPBI E.%
\end{APACrefauthors}%
\unskip\
\newblock
\APACrefYearMonthDay{1997}{}{}.
\newblock
{\BBOQ}\APACrefatitle {A procedure for optimizing tactical response in oil
  spill clean up operations} {A procedure for optimizing tactical response in
  oil spill clean up operations}.{\BBCQ}
\newblock
\APACjournalVolNumPages{European Journal of Operational
  Research}{102}{3}{554--574}.
\PrintBackRefs{\CurrentBib}

\bibitem [\protect \citeauthoryear {%
Stawicki%
\ \protect \BOthers {.}}{%
Stawicki%
\ \protect \BOthers {.}}{%
{\protect \APACyear {2020}}%
}]{%
stawicki20202019}
\APACinsertmetastar {%
stawicki20202019}%
\begin{APACrefauthors}%
Stawicki, S\BPBI P.%
, Jeanmonod, R.%
, Miller, A\BPBI C.%
, Paladino, L.%
, Gaieski, D\BPBI F.%
, Yaffee, A\BPBI Q.%
\BDBL {}Garg, M.%
\end{APACrefauthors}%
\unskip\
\newblock
\APACrefYearMonthDay{2020}{}{}.
\newblock
{\BBOQ}\APACrefatitle {The 2019--2020 novel coronavirus (severe acute
  respiratory syndrome coronavirus 2) pandemic: A joint {A}merican college of
  academic international medicine-world academic council of emergency medicine
  multidisciplinary {COVID-19} working group consensus paper} {The 2019--2020
  novel coronavirus (severe acute respiratory syndrome coronavirus 2) pandemic:
  A joint {A}merican college of academic international medicine-world academic
  council of emergency medicine multidisciplinary {COVID-19} working group
  consensus paper}.{\BBCQ}
\newblock
\APACjournalVolNumPages{Journal of Global Infectious Diseases}{12}{2}{47}.
\PrintBackRefs{\CurrentBib}

\bibitem [\protect \citeauthoryear {%
Strikholm%
}{%
Strikholm%
}{%
{\protect \APACyear {2006}}%
}]{%
strikholm2006determining}
\APACinsertmetastar {%
strikholm2006determining}%
\begin{APACrefauthors}%
Strikholm, B.%
\end{APACrefauthors}%
\unskip\
\newblock
\APACrefYearMonthDay{2006}{}{}.
\newblock
\APACrefbtitle {Determining the number of breaks in a piecewise linear
  regression model} {Determining the number of breaks in a piecewise linear
  regression model}\ \APACbVolEdTR {}{{SSE/EFI} Working Paper Series in
  Economics and Finance\ \BNUM~648}.
\newblock
\APACaddressInstitution{}{Stockholm School of Economics}.
\newblock
\begin{APACrefURL} \url{https://EconPapers.repec.org/RePEc:hhs:hastef:0648}
  \end{APACrefURL}
\PrintBackRefs{\CurrentBib}

\bibitem [\protect \citeauthoryear {%
Suits%
, Mason%
\BCBL {}\ \BBA {} Chan%
}{%
Suits%
\ \protect \BOthers {.}}{%
{\protect \APACyear {1978}}%
}]{%
suits1978spline}
\APACinsertmetastar {%
suits1978spline}%
\begin{APACrefauthors}%
Suits, D\BPBI B.%
, Mason, A.%
\BCBL {}\ \BBA {} Chan, L.%
\end{APACrefauthors}%
\unskip\
\newblock
\APACrefYearMonthDay{1978}{}{}.
\newblock
{\BBOQ}\APACrefatitle {Spline functions fitted by standard regression methods}
  {Spline functions fitted by standard regression methods}.{\BBCQ}
\newblock
\APACjournalVolNumPages{The {R}eview of {E}conomics and
  {S}tatistics}{}{}{132--139}.
\PrintBackRefs{\CurrentBib}

\bibitem [\protect \citeauthoryear {%
Sun%
, DePuy%
\BCBL {}\ \BBA {} Evans%
}{%
Sun%
\ \protect \BOthers {.}}{%
{\protect \APACyear {2014}}%
}]{%
sun2014multi}
\APACinsertmetastar {%
sun2014multi}%
\begin{APACrefauthors}%
Sun, L.%
, DePuy, G\BPBI W.%
\BCBL {}\ \BBA {} Evans, G\BPBI W.%
\end{APACrefauthors}%
\unskip\
\newblock
\APACrefYearMonthDay{2014}{}{}.
\newblock
{\BBOQ}\APACrefatitle {Multi-objective optimization models for patient
  allocation during a pandemic influenza outbreak} {Multi-objective
  optimization models for patient allocation during a pandemic influenza
  outbreak}.{\BBCQ}
\newblock
\APACjournalVolNumPages{Computers \& {O}perations {R}esearch}{51}{}{350--359}.
\PrintBackRefs{\CurrentBib}

\bibitem [\protect \citeauthoryear {%
Tomar%
\ \BBA {} Gupta%
}{%
Tomar%
\ \BBA {} Gupta%
}{%
{\protect \APACyear {2020}}%
}]{%
tomar2020prediction}
\APACinsertmetastar {%
tomar2020prediction}%
\begin{APACrefauthors}%
Tomar, A.%
\BCBT {}\ \BBA {} Gupta, N.%
\end{APACrefauthors}%
\unskip\
\newblock
\APACrefYearMonthDay{2020}{}{}.
\newblock
{\BBOQ}\APACrefatitle {Prediction for the spread of {COVID-19} in {India} and
  effectiveness of preventive measures} {Prediction for the spread of
  {COVID-19} in {India} and effectiveness of preventive measures}.{\BBCQ}
\newblock
\APACjournalVolNumPages{Science of {t}he {T}otal {E}nvironment}{728}{}{138762}.
\PrintBackRefs{\CurrentBib}

\bibitem [\protect \citeauthoryear {%
Van~Griensven%
\ \protect \BOthers {.}}{%
Van~Griensven%
\ \protect \BOthers {.}}{%
{\protect \APACyear {2016}}%
}]{%
van2016evaluation}
\APACinsertmetastar {%
van2016evaluation}%
\begin{APACrefauthors}%
Van~Griensven, J.%
, Edwards, T.%
, de Lamballerie, X.%
, Semple, M\BPBI G.%
, Gallian, P.%
, Baize, S.%
\BDBL {}Haba, N\BPBI Y.%
\end{APACrefauthors}%
\unskip\
\newblock
\APACrefYearMonthDay{2016}{}{}.
\newblock
{\BBOQ}\APACrefatitle {Evaluation of convalescent plasma for {E}bola virus
  disease in {G}uinea} {Evaluation of convalescent plasma for {E}bola virus
  disease in {G}uinea}.{\BBCQ}
\newblock
\APACjournalVolNumPages{New {E}ngland {J}ournal of {M}edicine}{374}{1}{33--42}.
\PrintBackRefs{\CurrentBib}

\bibitem [\protect \citeauthoryear {%
Velavan%
\ \BBA {} Meyer%
}{%
Velavan%
\ \BBA {} Meyer%
}{%
{\protect \APACyear {2020}}%
}]{%
velavan2020covid}
\APACinsertmetastar {%
velavan2020covid}%
\begin{APACrefauthors}%
Velavan, T\BPBI P.%
\BCBT {}\ \BBA {} Meyer, C\BPBI G.%
\end{APACrefauthors}%
\unskip\
\newblock
\APACrefYearMonthDay{2020}{}{}.
\newblock
{\BBOQ}\APACrefatitle {The {COVID-19} epidemic} {The {COVID-19}
  epidemic}.{\BBCQ}
\newblock
\APACjournalVolNumPages{Tropical {M}edicine \& {I}nternational
  {H}ealth}{25}{3}{278}.
\PrintBackRefs{\CurrentBib}

\bibitem [\protect \citeauthoryear {%
Weissman%
\ \protect \BOthers {.}}{%
Weissman%
\ \protect \BOthers {.}}{%
{\protect \APACyear {2020}}%
}]{%
weissman2020locally}
\APACinsertmetastar {%
weissman2020locally}%
\begin{APACrefauthors}%
Weissman, G\BPBI E.%
, Crane-Droesch, A.%
, Chivers, C.%
, Luong, T.%
, Hanish, A.%
, Levy, M\BPBI Z.%
\BDBL {}Halpern, S\BPBI D.%
\end{APACrefauthors}%
\unskip\
\newblock
\APACrefYearMonthDay{2020}{}{}.
\newblock
{\BBOQ}\APACrefatitle {Locally informed simulation to predict hospital capacity
  needs during the {COVID-19} pandemic} {Locally informed simulation to predict
  hospital capacity needs during the {COVID-19} pandemic}.{\BBCQ}
\newblock
\APACjournalVolNumPages{Annals of {I}nternal {M}edicine}{173}{1}{21--28}.
\PrintBackRefs{\CurrentBib}

\bibitem [\protect \citeauthoryear {%
Wilhelm%
\ \BBA {} Srinivasa%
}{%
Wilhelm%
\ \BBA {} Srinivasa%
}{%
{\protect \APACyear {1997}}%
}]{%
wilhelm1997prescribing}
\APACinsertmetastar {%
wilhelm1997prescribing}%
\begin{APACrefauthors}%
Wilhelm, W\BPBI E.%
\BCBT {}\ \BBA {} Srinivasa, A\BPBI V.%
\end{APACrefauthors}%
\unskip\
\newblock
\APACrefYearMonthDay{1997}{}{}.
\newblock
{\BBOQ}\APACrefatitle {Prescribing tactical response for oil spill clean up
  operations} {Prescribing tactical response for oil spill clean up
  operations}.{\BBCQ}
\newblock
\APACjournalVolNumPages{Management Science}{43}{3}{386--402}.
\PrintBackRefs{\CurrentBib}

\bibitem [\protect \citeauthoryear {%
Yang%
, Liu%
, Tsoka%
\BCBL {}\ \BBA {} Papageorgiou%
}{%
Yang%
\ \protect \BOthers {.}}{%
{\protect \APACyear {2016}}%
}]{%
yang2016mathematical}
\APACinsertmetastar {%
yang2016mathematical}%
\begin{APACrefauthors}%
Yang, L.%
, Liu, S.%
, Tsoka, S.%
\BCBL {}\ \BBA {} Papageorgiou, L\BPBI G.%
\end{APACrefauthors}%
\unskip\
\newblock
\APACrefYearMonthDay{2016}{}{}.
\newblock
{\BBOQ}\APACrefatitle {Mathematical programming for piecewise linear regression
  analysis} {Mathematical programming for piecewise linear regression
  analysis}.{\BBCQ}
\newblock
\APACjournalVolNumPages{Expert {S}ystems with {A}pplications}{44}{}{156--167}.
\PrintBackRefs{\CurrentBib}

\bibitem [\protect \citeauthoryear {%
Yao%
, Yang%
\BCBL {}\ \BBA {} Zhan%
}{%
Yao%
\ \protect \BOthers {.}}{%
{\protect \APACyear {2013}}%
}]{%
yao2013novel}
\APACinsertmetastar {%
yao2013novel}%
\begin{APACrefauthors}%
Yao, D.%
, Yang, J.%
\BCBL {}\ \BBA {} Zhan, X.%
\end{APACrefauthors}%
\unskip\
\newblock
\APACrefYearMonthDay{2013}{}{}.
\newblock
{\BBOQ}\APACrefatitle {A novel method for disease prediction: {H}ybrid of
  random forest and multivariate adaptive regression splines} {A novel method
  for disease prediction: {H}ybrid of random forest and multivariate adaptive
  regression splines}.{\BBCQ}
\newblock
\APACjournalVolNumPages{Journal of {C}omputers}{8}{1}{170--177}.
\PrintBackRefs{\CurrentBib}

\bibitem [\protect \citeauthoryear {%
Yarmand%
, Ivy%
, Denton%
\BCBL {}\ \BBA {} Lloyd%
}{%
Yarmand%
\ \protect \BOthers {.}}{%
{\protect \APACyear {2014}}%
}]{%
yarmand2014optimal}
\APACinsertmetastar {%
yarmand2014optimal}%
\begin{APACrefauthors}%
Yarmand, H.%
, Ivy, J\BPBI S.%
, Denton, B.%
\BCBL {}\ \BBA {} Lloyd, A\BPBI L.%
\end{APACrefauthors}%
\unskip\
\newblock
\APACrefYearMonthDay{2014}{}{}.
\newblock
{\BBOQ}\APACrefatitle {Optimal two-phase vaccine allocation to geographically
  different regions under uncertainty} {Optimal two-phase vaccine allocation to
  geographically different regions under uncertainty}.{\BBCQ}
\newblock
\APACjournalVolNumPages{European {J}ournal of {O}perational
  {R}esearch}{233}{1}{208--219}.
\PrintBackRefs{\CurrentBib}

\bibitem [\protect \citeauthoryear {%
Yin%
\ \BBA {} B{\"u}y{\"u}ktahtak{\i}n%
}{%
Yin%
\ \BBA {} B{\"u}y{\"u}ktahtak{\i}n%
}{%
{\protect \APACyear {2021}}%
}]{%
yin2021multi}
\APACinsertmetastar {%
yin2021multi}%
\begin{APACrefauthors}%
Yin, X.%
\BCBT {}\ \BBA {} B{\"u}y{\"u}ktahtak{\i}n, {\.I}\BPBI E.%
\end{APACrefauthors}%
\unskip\
\newblock
\APACrefYearMonthDay{2021}{}{}.
\newblock
{\BBOQ}\APACrefatitle {A multi-stage stochastic programming approach to
  epidemic resource allocation with equity considerations} {A multi-stage
  stochastic programming approach to epidemic resource allocation with equity
  considerations}.{\BBCQ}
\newblock
\APACjournalVolNumPages{Health Care Management Science}{24}{3}{597--622}.
\PrintBackRefs{\CurrentBib}

\bibitem [\protect \citeauthoryear {%
Zhou%
, Zhong%
\BCBL {}\ \BBA {} Guan%
}{%
Zhou%
\ \protect \BOthers {.}}{%
{\protect \APACyear {2007}}%
}]{%
zhou2007treatment}
\APACinsertmetastar {%
zhou2007treatment}%
\begin{APACrefauthors}%
Zhou, B.%
, Zhong, N.%
\BCBL {}\ \BBA {} Guan, Y.%
\end{APACrefauthors}%
\unskip\
\newblock
\APACrefYearMonthDay{2007}{}{}.
\newblock
{\BBOQ}\APACrefatitle {Treatment with convalescent plasma for influenza {A}
  ({H5N1}) infection} {Treatment with convalescent plasma for influenza {A}
  ({H5N1}) infection}.{\BBCQ}
\newblock
\APACjournalVolNumPages{New {E}ngland {J}ournal of
  {M}edicine}{357}{14}{1450--1451}.
\PrintBackRefs{\CurrentBib}

\bibitem [\protect \citeauthoryear {%
Zietz%
, Zucker%
\BCBL {}\ \BBA {} Tatonetti%
}{%
Zietz%
\ \protect \BOthers {.}}{%
{\protect \APACyear {2020}}%
}]{%
zietz2020associations}
\APACinsertmetastar {%
zietz2020associations}%
\begin{APACrefauthors}%
Zietz, M.%
, Zucker, J.%
\BCBL {}\ \BBA {} Tatonetti, N\BPBI P.%
\end{APACrefauthors}%
\unskip\
\newblock
\APACrefYearMonthDay{2020}{}{}.
\newblock
{\BBOQ}\APACrefatitle {Associations between blood type and {COVID-19}
  infection, intubation, and death} {Associations between blood type and
  {COVID-19} infection, intubation, and death}.{\BBCQ}
\newblock
\APACjournalVolNumPages{Nature {C}ommunications}{11}{1}{1--6}.
\PrintBackRefs{\CurrentBib}

\end{thebibliography}

\end{document}